\documentclass[a4paper,11pt]{article}
\pdfoutput=1

\usepackage[english]{babel}
\usepackage{jheppub}
\usepackage[T1]{fontenc}
\usepackage{graphicx,calc}
\usepackage{epsfig}
\usepackage{amsmath}
\usepackage{amssymb}
\usepackage{cleveref}
\usepackage{float}
\usepackage{placeins}
\usepackage{braket}
\usepackage{ifthen}
\usepackage{slashed}
\usepackage{mathdots}
\usepackage{feynmf}
\usepackage{color}
\usepackage{multicol}
\usepackage{caption}
\usepackage{subcaption}
\usepackage{array}
\usepackage[export]{adjustbox}
\usepackage{subcaption}
\usepackage{multicol}
\usepackage{comment}
\usepackage{colortbl}
\usepackage{xcolor}
\usepackage{booktabs}

\newcommand{\eps}{\epsilon}
\newcommand{\ii}{{\mathrm{i}}}
\newcommand{\dd}{{\mathrm{d}}}
\newcommand{\vMJ}{\vec{\mathcal{J}}}
\newcommand{\dlog}{\dd \log}
\newcommand{\zt}{\tilde{z}}
\newcommand{\ct}{\tilde{c}}
\newcommand{\cC}{\mathcal{C}}
\newcommand{\cP}{\mathcal{P}}

\newcommand{\brk}[1]{(#1)}
\newcommand{\bigbrk}[1]{\bigl(#1\bigr)}

\newcommand{\sbrk}[1]{[#1]}
\newcommand{\Bigsbrk}[1]{\Bigl[#1\Bigr]}
\newcommand{\brc}[1]{\{#1\}}
\newcommand{\Bigbrc}[1]{\Bigl\{#1\Bigr\}}
\newcommand{\supbrk}[1]{^{\brk{#1}}}

\newcommand{\code}[1]{\texttt{#1}}
\newcommand{\soft}[1]{\textsc{#1}}

\allowdisplaybreaks

\title{Two-loop Feynman integrals for leading colour $t\bar{t}W$ production at hadron colliders}

\author[a]{Matteo Becchetti,}
\author[a]{Dhimiter Canko,}
\author[a]{Vsevolod Chestnov,}
\author[a]{Tiziano Peraro,}
\author[a]{Mattia Pozzoli,}
\author[b]{Simone Zoia}

\affiliation[a]{Dipartimento di Fisica e Astronomia, Universit\`{a}	 di Bologna, \\
INFN, Sezione di Bologna, \\
via Irnerio 46, I-40126 Bologna, Italy}
\affiliation[b]{Physik-Institut, Universit\"{a}t Z\"{u}rich, \\
Winterthurerstrasse 190, CH-8057 Z\"{u}rich, Switzerland}

\emailAdd{matteo.becchetti@unibo.it}
\emailAdd{dhimiter.canko2@unibo.it}
\emailAdd{vsevolod.chestnov@unibo.it}
\emailAdd{tiziano.peraro@unibo.it}
\emailAdd{mattia.pozzoli@unibo.it}
\emailAdd{simone.zoia@physik.uzh.ch}

\preprint{ZU-TH 28/25}

\abstract{
We compute a complete set of the two-loop Feynman integrals that are required for the next-to-next-to-leading order QCD corrections to on-shell top-pair production in association with a $W$ boson at hadron colliders in the leading colour approximation.
These Feynman integrals also contribute to Higgs or $Z$-boson production in association with a top pair.
We employ the method of differential equations (DEs), facilitated by the use of finite field methods to handle the algebraic complexity stemming from the seven-scale kinematics.
The presence of the top quark in the virtual propagators, in addition to the mass of the external $W$ boson, gives rise to nested square roots and three elliptic curves.
We obtain DEs that depend at most quadratically on the dimensional regulator $\eps$ for sectors where these analytic structures appear, and are $\eps$-factorised otherwise.
We express the DEs in terms of a minimal set of differential one-forms, separating the logarithmic ones.
We solve the DEs numerically in the physical kinematic region, with the method of generalised power series expansions.}

\begin{document}
\maketitle
\flushbottom

\section{Introduction}
\label{Introduction}

Feynman integrals are a cornerstone of theoretical particle physics, playing a vital role in obtaining precise predictions in quantum field theory.
This field lies at the intersection of several disciplines, bridging particle physics, mathematics, string theory, and more recently finding application also in gravitational wave physics and cosmology.
The main motivation for our work stems from the precision physics programme of the Large Hadron Collider (LHC).
In order for us to exploit fully its data, theoretical predictions for a wide range of scattering processes must reach at least the next-to-next-to-leading order (NNLO) in QCD.
The current frontier for NNLO QCD calculations are $2 \to 3$ processes~\cite{Andersen:2024czj,Huss:2025nlt}, for which the main bottleneck is the computation of the required two-loop five-particle scattering amplitudes and Feynman integrals.
Driven by this phenomenological motivation, significant progress has been made in recent years in the computation of both Feynman integrals and scattering amplitudes, resulting
in an increasing availability of NNLO QCD predictions for high-multiplicity processes.

This progress has been driven by three main developments.
Firstly, the advent of \emph{finite-field} techniques~\cite{vonManteuffel:2014ixa,Peraro:2016wsq} has drastically mitigated the appearance of large intermediate expressions that naturally arise in these computations, a problem that is worsened by the large number of kinematic variables in $2\to 3$ processes.
Secondly, substantial improvements in \emph{integration-by-part} (IBP)~\cite{Tkachov:1981wb,Chetyrkin:1981qh,Laporta:2000dsw} reduction methods~\cite{Gluza:2010ws,Ita:2015tya,Larsen:2015ped,Wu:2023upw,Guan:2024byi,Chestnov:2024mnw} have mitigated what is one of the main bottlenecks in these computations.
Finally, the deepened understanding of the relevant special functions~\cite{Abreu:2022mfk} has led to a systematic methodology to write the Feynman integrals in terms of \emph{bases of special functions} that can be evaluated numerically efficiently and unlock simplifications at the amplitude level~\cite{Gehrmann:2018yef,Chicherin:2020oor,Chicherin:2021dyp,Abreu:2023rco,Gehrmann:2024tds}.
The application of these powerful mathematical techniques requires that a certain \emph{canonical form}~\cite{Henn:2013pwa} is obtained for the
\emph{differential equations} (DEs)~\cite{Barucchi:1973zm,Kotikov:1990kg,Kotikov:1991hm,Gehrmann:1999as,Bern:1993kr,Henn:2013pwa} governing the Feynman integrals.

Crucially, the vast majority of the NNLO QCD results do not involve scattering processes with massive virtual particles.
The latter are considerably more difficult, owing to an increased algebraic complexity and to the appearance of special functions for which the mathematical formalism is not yet suitable for phenomenology.
Processes of this type are of great interest for LHC phenomenology~\cite{Andersen:2024czj,Huss:2025nlt};
key examples are the production of a top-quark pair in association with either a jet ($t\bar{t}j$), an electroweak boson $W$/$Z$ ($t\bar{t}W/Z$), or a Higgs boson ($t\bar{t}H$).
For $t\bar{t}j$ production, a complete set of two-loop Feynman integrals in the leading colour approximation have been recently computed~\cite{Badger:2022hno,Badger:2024fgb,Becchetti:2025oyb},
enabling a first numerical evaluation of the two-loop finite remainders in the gluon channel~\cite{Badger:2024dxo}.
For $t\bar{t}H$ production, a class of two-loop Feynman integrals related to diagrams with
light-quark loops have been studied in ref.~\cite{FebresCordero:2023pww}, and numerical results have been obtained for the $n_f$ part of the quark-initiated two-loop scattering amplitude~\cite{Agarwal:2024jyq}.
Analytic results for amplitudes of this type are available only at one loop, to the second order in the dimensional regulator, for $t\bar{t}j$~\cite{Badger:2022mrb}, $t\bar{t}H$~\cite{Buccioni:2023okz}, and $t\bar{t}W$~\cite{Becchetti:2025osw}.
NNLO QCD theoretical predictions are available for $t\bar{t}H$ and $t\bar{t}W$ only with approximated two-loop amplitudes~\cite{Catani:2022mfv,Buonocore:2023ljm,Devoto:2024nhl,Balsach:2025jcw}.

In this paper we study the two-loop Feynman integrals appearing in the NNLO QCD corrections to $t\bar{t}W$ production at hadron colliders in the leading colour approximation.
This process has one of the heaviest signatures at the LHC, it is relevant to several searches of physics beyond the Standard Model (SM)~\cite{Buckley:2015lku,Dror:2015nkp,BessidskaiaBylund:2016jvp}, and gives an important background for several other interesting SM processes, such as $t\bar{t}H$ and four-top production.
Most importantly, the experimental measurements~\cite{ATLAS:2015qtq,CMS:2015uvn,ATLAS:2016wgc,CMS:2017ugv,ATLAS:2019fwo,CMS:2022tkv,ATLAS:2024moy} are in tension with the SM predictions.
Experimental collaborations are currently comparing measurements against NLO QCD+EW predictions~\cite{Frederix:2021agh}.
The NLO QCD and electroweak (EW) corrections have been computed in refs.~\cite{Badger:2010mg,Campbell:2012dh,Maltoni:2015ena} and~\cite{Frixione:2015zaa,Frederix:2017wme}, respectively.
Refs.~\cite{Li:2014ula,Broggio:2016zgg,Kulesza:2018tqz,Broggio:2019ewu,Kidonakis:2023jpj} included soft-gluon effects,
while refs.~\cite{Bevilacqua:2020pzy,Denner:2020hgg,Bevilacqua:2020srb} and ref.~\cite{Denner:2021hqi} included the off-shell effects at NLO QCD and NLO QCD+EW, respectively.
The NLO QCD+EW predictions still have relatively large uncertainties, calling for the computation of the NNLO QCD corrections.
First steps towards this end were made in ref.~\cite{Buonocore:2023ljm}, where the NNLO~QCD~+~NLO~EW corrections to the inclusive cross section have been computed by using approximations for the finite part of the two-loop amplitudes~\cite{Catani:2022mfv,Barnreuther:2013qvf,Penin:2005eh,Mitov:2006xs,Becher:2007cu}.
On the one hand, obtaining exact NNLO QCD corrections remains a priority, both to confirm the robustness of the approximations and to analyse phase-space regions where such approximations are expected to be less accurate.
On the other, ref.~\cite{Buonocore:2023ljm} shows that the two-loop amplitudes are indeed the main bottleneck.

In view of removing this obstacle, we employ the method of DEs to tackle the computation of the two-loop five-particle Feynman integrals appearing in the leading colour approximation.
The presence of internal massive propagators entails a substantial leap in complexity with respect to the massless case.
Not only do the additional masses increase the algebraic complexity of the expressions, but they also bring about transcendental functions associated with higher-genus geometries~\cite{Bourjaily:2022bwx}, such as periods of elliptic curves.
The problem of extending the framework of canonical DEs to Feynman integrals with higher-genus geometries has attracted a lot of effort in recent years~\cite{Adams:2017tga,Adams:2018yfj,Frellesvig:2021hkr,Gorges:2023zgv,Dlapa:2022wdu,Pogel:2022ken,Pogel:2022vat,Frellesvig:2023iwr,Driesse:2024feo,Duhr:2024uid,Duhr:2025lbz,Chen:2025hzq}.
Yet, the application of these methods is mostly confined to processes with comparatively simpler $2\to 2$ kinematics~\cite{Duhr:2021fhk,Muller:2022gec,Gorges:2023zgv,Delto:2023kqv,Duhr:2024bzt,Schwanemann:2024kbg,Becchetti:2025rrz}, with a single recent exception for a family of Feynman integrals relevant to $t\bar{t}j$ production involving an elliptic curve~\cite{Becchetti:2025oyb}.
Moreover, although promising approaches based on series expansions are being explored~\cite{Walden:2020odh,Badger:2022mrb,Delto:2023kqv,Duhr:2024bzt,Forner:2024ojj,Becchetti:2025rrz}, evaluating efficiently the solution to canonical DEs containing elliptic functions is still an open problem, especially for processes with three or more particles in the final state.
The Feynman integrals considered in this work involve three (non-isomorphic) \emph{elliptic curves} (see \cref{sec:Elliptic_curves}).
While two of them have four-point kinematics and are similar to the curve encountered in $t\bar{t}j$ production~\cite{Badger:2024fgb,Becchetti:2025oyb},
the third curve has the full five-particle kinematics and is significantly more complicated.
Moreover, we also encounter \emph{nested square roots} (see \cref{sec:Nested_square_roots}), a feature for which there are still scarce data~\cite{FebresCordero:2023pww,Badger:2024fgb,Becchetti:2025oyb}.
Given the extreme computational challenge posed by the five-point elliptic integrals, and the fact that the efficient numerical evaluation would remain an open problem even if we achieved canonical DEs for them, we aim to construct systems of DEs that satisfy milder constraints, without introducing elliptic functions and nested square roots, as done in ref.~\cite{Badger:2024fgb,Badger:2024dxo}.
As shown in ref.~\cite{Badger:2024dxo}, such systems of DEs can be used to set up an efficient computation of the two-loop amplitude.
Moreover, the absence of elliptic functions allows us to use the method of generalised power series~\cite{Moriello:2019yhu} to evaluate numerically the solution with publicly available tools~\cite{Hidding:2020ytt,Armadillo:2022ugh,Prisco:2025wqs}.

The paper is structured as follows.
In \cref{Kinematics_and_Families} we define the families of Feynman integrals and introduce the notation.
In \cref{sec:Bases_and_DEs} we detail the construction of the integral bases and of the DEs they obey, and discuss the most interesting analytic structures: the elliptic curves and the nested square roots.
We describe how we represent the DEs and how we solve them numerically in \cref{Results}.
We draw our conclusions and give an outlook on future work in \cref{Conclusions}.
We also provide a number of appendices: in \cref{Ancillary} we describe our ancillary files~\cite{zenodo}, in \cref{New_Sectors} we display the five-point master integrals that we did not discuss in \cref{sec:Bases_and_DEs}, and in \cref{app:elliptic_details} we give more technical details on the elliptic curve of \cref{sec:five-point_elliptic}.

\section{Notation and definitions}
\label{Kinematics_and_Families}

We study the production of a $W$ boson in association with two top quarks at hadron colliders.
We set the top quarks on their mass shell, and leave the $W$ boson off shell.
We denote by $p_1$ and $p_2$ the momenta of the top quarks, by $p_4$ the momentum of the $W$ boson, and by $p_3$ and $p_5$ the momenta of the light quarks.
The momenta are all outgoing, and satisfy momentum conservation,
\begin{align}
p_1 + p_2 + p_3 + p_4 + p_5 = 0 \,,
\end{align}
as well as the following on-shellness conditions,
\begin{align}
p_1^2 = p_2^2 = m_t^2 \,, \qquad \qquad p_3^2=p_5^2=0 \,,
\end{align}
where $m_t$ is the mass of the top quark.
We parametrise the kinematics by seven scalar Lorentz-invariant quantities, which we choose as
\begin{equation}
\label{eq:invariants}
\vec{x} = \left\{s_{12},s_{23},s_{34},s_{45},s_{15},m_t^2,m_w^2 \right\} \,,
\end{equation}
where $s_{ij} = (p_i + p_j)^2$, and $m_w^2 = p_4^2$ is the off-shellness of the $W$ boson.

\begin{figure}[t]
\centering
\begin{subfigure}[b]{0.497\textwidth}
\centering
\includegraphics[scale=0.36]{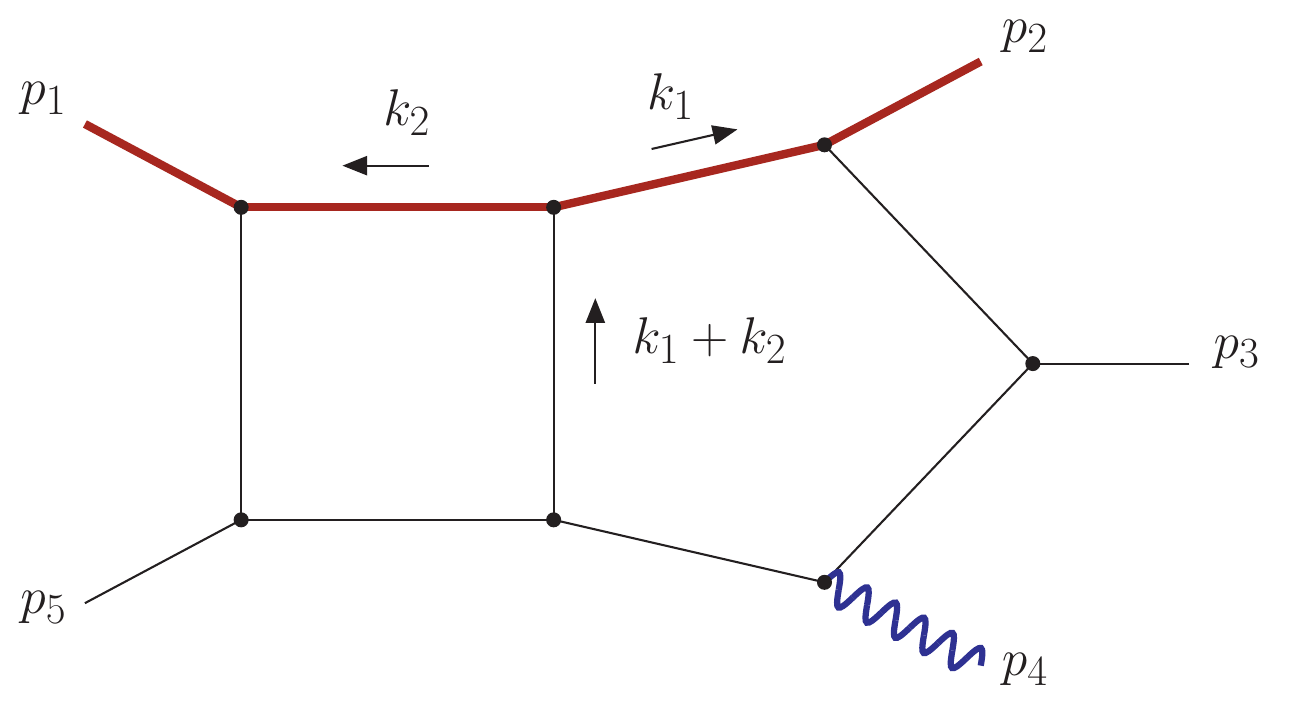}
\caption{Family $F_1$}
\end{subfigure}
\hfill
\begin{subfigure}[b]{0.496\textwidth}
\centering
\includegraphics[scale=0.36]{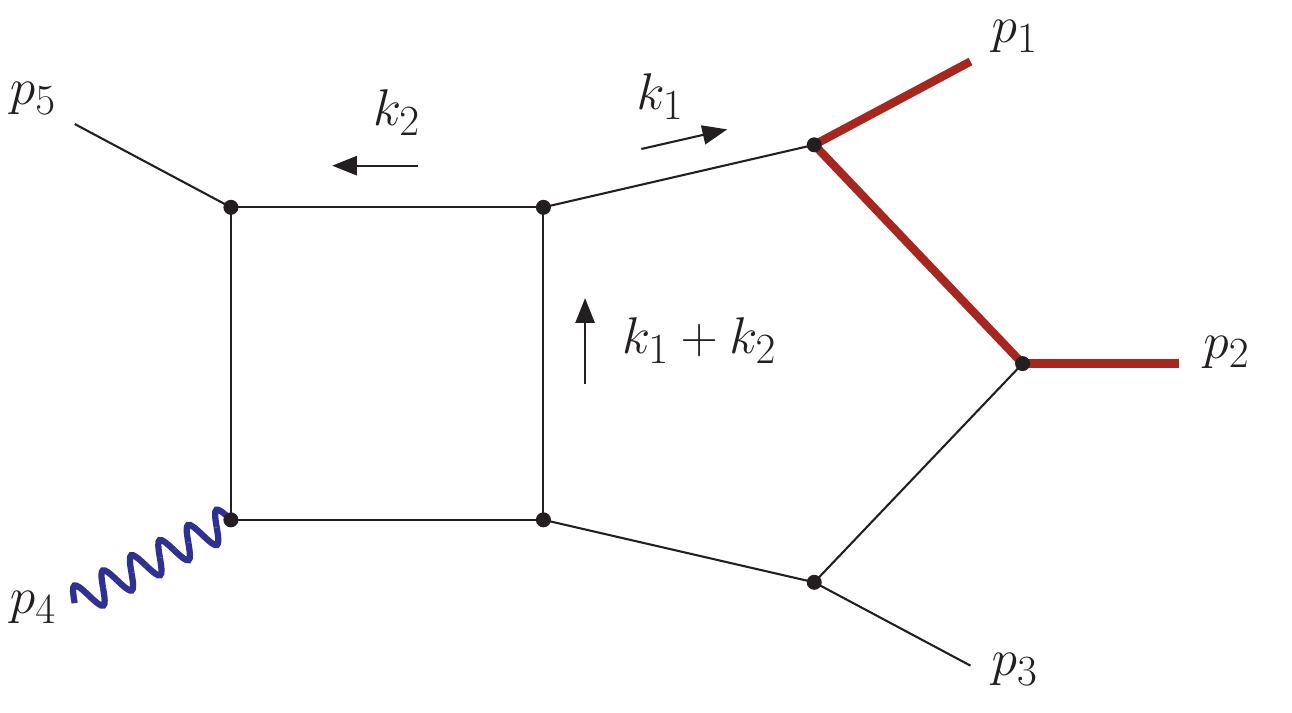}
\caption{Family $F_2$}
\end{subfigure}
\hfill
\begin{subfigure}[b]{0.496\textwidth}
\centering
\includegraphics[scale=0.36]{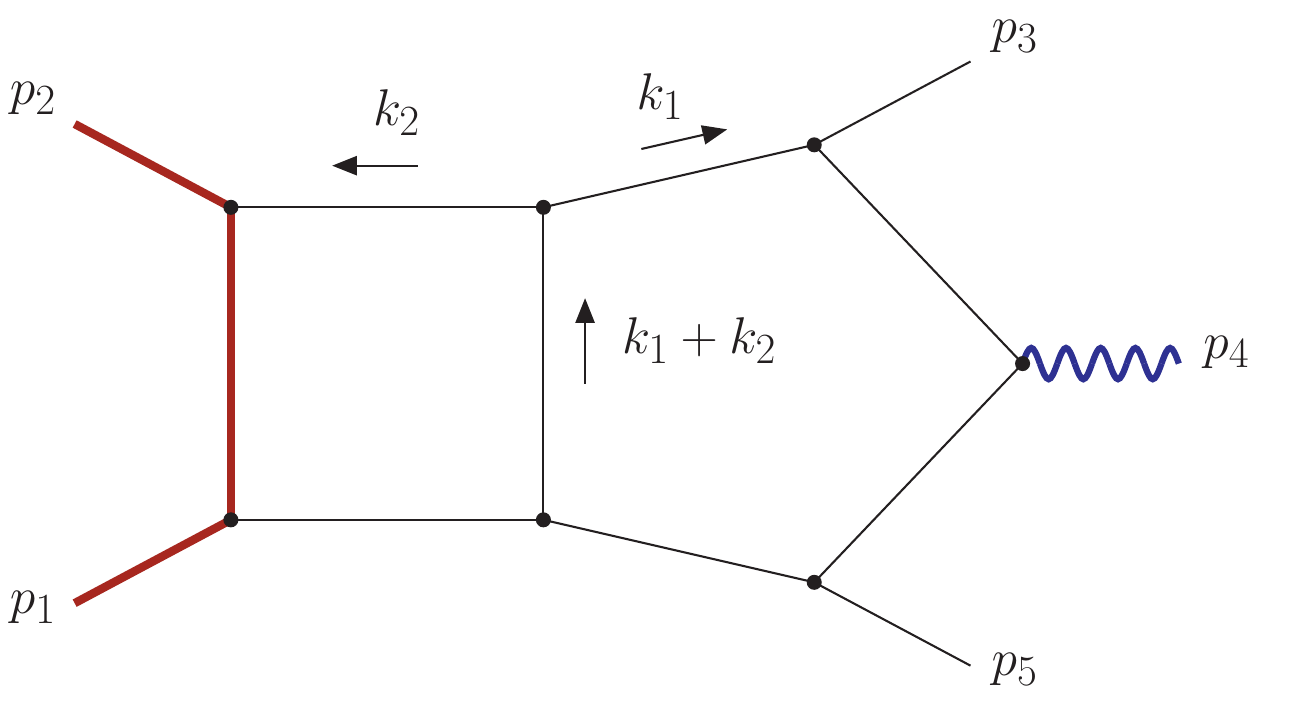}
\caption{Family $F_3$}
\end{subfigure}
\caption{The two-loop integral families contributing to $t \bar{t} W$ production at hadron colliders in the leading colour approximation. The red lines correspond to (anti-)top quarks, the blue curved line to the $W$ boson, and the black lines to massless particles.}
\label{fig:Families}
\end{figure}

In the leading colour approximation, all Feynman integrals appearing in the two-loop amplitude can be reduced to integrals of the three two-loop five-point families shown in \cref{fig:Families}, or to products of the one-loop integrals computed in ref.~\cite{Becchetti:2025osw}.
We label the families in \cref{fig:Families} by $F_1$, $F_2$ and $F_3$,
and define the scalar integrals associated with family $F \in \left\{F_1, F_2, F_3 \right\}$~as
\begin{align}
\label{eq:famdef}
I^{(F)}_{\vec{\nu}}\left[N\right]
& =\int  \frac{\dd^{d}k_1\,e^{\epsilon\gamma_E}}{\ii \pi^{\frac{d}{2}}} \, \frac{\dd^{d}k_2\,e^{\epsilon\gamma_E}}{\ii \pi^{\frac{d}{2}}} \,
\frac{N}
{\prod_{i=1}^{11} D^{\nu_i}_{F,i} }\,, \\
I^{(F)}_{\vec{\nu}}
& = I^{(F)}_{\vec{\nu}}\left[1\right] \,,
\end{align}
where $\vec{\nu} = (\nu_1, \cdots, \nu_{11}) \in \mathbb{Z}^{11}$, $d = 4 - 2 \eps$ is the space-time dimension, and $N$ is a polynomial in the scalar products $k_i \cdot p_j$ and $k_i \cdot k_j$.
The inverse propagators $D_{F,i}$ are defined in \cref{tab:propagators}.
$D^{\nu_9}_{F,9}$, $D^{\nu_9}_{F,10}$ and $D^{\nu_9}_{F,11}$ are irreducible scalar products, i.e., $\nu_9, \nu_{10}, \nu_{11} \le 0$.
We omit the dependence on $\vec{x}$ and $\eps$.

\begin{table}[t]
\centering
\begin{tabular}{|c|c|c|c|}
\hline
& $F_1$ & $F_2$ & $F_3$ \\
\hline
$D_{F,1}$ & $k_1^2-m_t^2$ & $k_1^2$ & $k_1^2$ \\
$D_{F,2}$ & $(k_1-p_2)^2$ & $(k_1-p_1)^2-m_t^2$ & $(k_1-p_3)^2$ \\
$D_{F,3}$ & $(k_1-p_{23})^2$ & $(k_1-p_{12})^2$ & $(k_1-p_{34})^2$ \\
$D_{F,4}$ & $(k_1-p_{234})^2$ & $(k_1-p_{123})^2$ & $(k_1+p_{12})^2$ \\
$D_{F,5}$ & $k_2^2-m_t^2$ & $k_2^2$  & $k_2^2$ \\
$D_{F,6}$ & $(k_2-p_1)^2$ & $(k_2+p_{1234})^2$ & $(k_2-p_2)^2-m_t^2$ \\
$D_{F,7}$ & $(k_2+p_{234})^2$ & $(k_2+p_{123})^2$ & $(k_2-p_{12})^2$ \\
$D_{F,8}$ & $(k_1+k_2)^2$ & $(k_1+k_2)^2$ & $(k_1+k_2)^2$ \\
$D_{F,9}$ & $(k_1+p_1)^2-m_t^2$ & $(k_1-p_{1234})^2$ & $(k_1+p_2)^2-m_t^2$ \\
$D_{F,10}$ & $(k_2+p_2)^2-m_t^2$ & $(k_2+p_1)^2-m_t^2$ & $(k_2+p_3)^2$ \\
$D_{F,11}$ & $(k_2+p_{23})^2-m_t^2$ & $(k_2+p_{12})^2$ & $(k_2+p_{34})^2$ \\
\hline
\end{tabular}
\caption{Inverse propagators $D_{F,i}$ in \cref{eq:famdef} for the integral families drawn in \cref{fig:Families}. We use the shorthand $p_{i\cdots l} = p_i+\cdots + p_l$.}
\label{tab:propagators}
\end{table}

It is useful to introduce Gram determinants,
\begin{equation}
\label{eq:gram-definitions}
\begin{split}
{\rm G}\begin{pmatrix} a_1,\ldots,a_n \\ b_1,\ldots,b_n \end{pmatrix} & = {\text{det}}\left( 2\, a_i\cdot b_j \right) \,, \\
{\rm G} (a_1,\ldots,a_n ) & = {\rm G}\begin{pmatrix} a_1,\ldots,a_n \\ a_1,\ldots,a_n \end{pmatrix} \,.
\end{split}
\end{equation}
They play a role in describing the singularity structure of Feynman integrals, constructing master integrals (MIs), and defining the scattering regions.
For example, we anticipate that the MIs constructed in \cref{sec:Bases_and_DEs} contain the following square roots:

\begin{align}
\label{eq:roots1}
\begin{aligned}
r_1&= \sqrt{{\rm G}(p_1,p_2,p_3,p_4)} \,, \qquad \quad & r_2&=\sqrt{-{\rm G}(p_{34},p_1)} \,, \\
r_3&=\sqrt{-{\rm G}(p_{13},p_4)} \,, & r_4&=\sqrt{-{\rm G}(p_{45},p_2)} \,, \\
r_5&=\sqrt{-{\rm G}(p_{45},p_1)} \,, & r_6 &=\sqrt{-{\rm G}(p_{15},p_4)} \,, \\
r_7 & = \sqrt{-{\rm G}(p_{34},p_2)} \,,  & r_8 & =\sqrt{-{\rm G}(p_{12},p_4)} \,, \\
r_9 & =\sqrt{-{\rm G}(p_1,p_2)} \,, & r_{10} & =\sqrt{-{\rm G}(p_2,p_4)} \,, \\
r_{11} & =\sqrt{-{\rm G}(p_1,p_4)} \,,
\end{aligned}
\end{align}
where $p_{ij} = p_i + p_j$.
The square roots $r_1$~--~$r_9$ appear already at one-loop order~\cite{Becchetti:2025osw}.\footnote{
The momenta ref.~\cite{Becchetti:2025osw} are to be relabelled as $(p_1,p_2,p_3,p_4,p_5) \to (p_3, p_5, p_1, p_2, p_4)$ in order to match our conventions.}
Out of these, $r_3$ does not appear in the families shown in \cref{fig:Families}; nonetheless we keep it in the list for consistency with the one-loop computation, and because it will appear in the two-loop amplitude as a permutation of $r_6$.
In addition, we have three square roots whose arguments are not Gram determinants:
\begin{align}
    \label{eq:roots2}
\begin{aligned}
& r_{12} =\sqrt{2 m_w^2 s_{45} {\rm G}\begin{pmatrix} p_1 , p_2 \\ p_{34}, p_{15} \end{pmatrix}  - m_w^4 {\rm G}(p_1, p_2) - s_{45}^2 {\rm G}(p_{34},p_{15})} \,, \\
& r_{13} =\sqrt{2 m_w^2 s_{45} {\rm G}\begin{pmatrix} p_1 , p_5 \\ p_{23}, p_{45} \end{pmatrix}  - m_w^4 {\rm G} (p_{23},p_{45}) - s_{45}^2{\rm G} (p_1,p_5) } \,,\\
& r_{14} =\sqrt{2 m_w^2 s_{34} {\rm G}\begin{pmatrix} p_1 , p_2 \\ p_{23}, p_{45} \end{pmatrix}  - m_w^4 {\rm G}(p_1, p_2) - s_{34}^2 {\rm G}(p_{45}, p_{23})} \,.
\end{aligned}
\end{align}
Here, we used Gram determinants merely to obtain a more compact representation.
We note that $r_{12}$ may be expressed as the square root of a Cayley determinant (see e.g.\ ref.~\cite{Badger:2024fgb}).

We use Gram determinants also to write the scalar products of the $(-2 \eps)$-dimensional components of the loop momenta in terms of scalar products of external and loop momenta:
\begin{align} \label{eq:muij}
\begin{aligned}
\mu_{ij} & = k_i^{[-2\eps]}\cdot k_j^{[-2\eps]} \\
& = \frac{{\rm G}\begin{pmatrix} k_i,p_1,p_2,p_3,p_4 \\ k_j,p_1,p_2,p_3,p_4 \end{pmatrix}}{ {\rm G}( p_1,p_2,p_3,p_4 )} \,,
\end{aligned}
\end{align}
where we decompose the loop momenta as $k_i = k_i^{[4]} + k_i^{[-2\eps]}$, with $k_i^{[4]} \cdot k_j^{[-2\eps]} = p_i \cdot k_j^{[-2\eps]} = 0$.
The insertion of $\mu_{ij}$'s in the numerators of Feynman integrals has been known for a long time to be useful
to define MIs that satisfy canonical DEs~\cite{Arkani-Hamed:2010zjl,Arkani-Hamed:2010pyv} (see also e.g.\ refs.~\cite{Gehrmann:2015bfy,Badger:2016ozq,Chicherin:2018old,Abreu:2020jxa,Badger:2024fgb,Abreu:2024yit}), and will play an important role in \cref{sec:Bases_and_DEs}.

\section{Construction of the integral bases}
\label{sec:Bases_and_DEs}

In this section, we discuss the construction of the bases of MIs. 
In order to identify a starting basis for each family, we used the private \soft{Mathematica} package \soft{FFIntRed} and \soft{NeatIBP}\footnote{We use a modified version of \soft{NeatIBP} that employs \soft{FiniteFlow} for the row reduction of linear systems. In our experience, this combination of tools proved to be more stable when dealing with the most challenging integral sectors.}~\cite{Wu:2023upw} to generate a system of IBP identities~\cite{Tkachov:1981wb,Chetyrkin:1981qh} and symmetry relations, and solved it with a variant of the Laporta algorithm~\cite{Laporta:2000dsw} in the finite-field framework \soft{FiniteFlow}~\cite{Peraro:2016wsq,Peraro:2019svx}. The number of MIs is 141, 122 and 131 for $F_1$, $F_2$ and $F_3$, respectively. Modding out by permutations of the external legs, there are 30 genuinely new sectors, containing 85 MIs which have not been computed previously.
For the sectors that are part of integral families that have been computed in the past~\cite{Henn:2014lfa, DiVita:2017xlr, Abreu:2020jxa, Long:2021vse, Badger:2022hno, FebresCordero:2023pww, Badger:2024fgb}, we use the MIs defined therein.

We denote by $\vMJ_F = \bigl(\mathrm{I}_1^{F},\mathrm{I}_2^{F}, \ldots \bigr)^{\top}$ the vector of MIs of family $F$.
We normalise the MIs so that their Laurent expansion around $\eps=0$ starts at order $\eps^0$.
For each family $F$, the vector $\vMJ_F$ satisfies a system of first-order linear partial DEs (alias a Pfaffian system)~\cite{Barucchi:1973zm,Kotikov:1990kg,Kotikov:1991hm,Gehrmann:1999as,Bern:1993kr} of the general form
\begin{align} \label{eq:DEsGen}
    \frac{\partial}{\partial x_a} \vMJ_F\brk{\vec{x}, \eps}
    =
    A_{x_a}\supbrk{F}\brk{\vec{x}, \eps}
    \cdot
    \vMJ_F\brk{\vec{x}, \eps}
    \>,
    \qquad
    \forall \ a = 1, \ldots, 7
    \> ,
\end{align}
where we recall that $\vec{x}$ denotes cumulatively the kinematic variables, cfr.~\cref{eq:invariants}.
The matrices $A_{x_a}\supbrk{F}\brk{\vec{x}, \eps}$ on the right-hand sides are computed by rewriting the derivatives of the MIs in terms of the MIs themselves through IBP identities.
It is beneficial to write the system in \cref{eq:DEsGen} compactly by using the exterior derivative
notation, as
\begin{align} \label{eq:DE_one_form}
    \dd \vMJ_F\brk{\vec{x}, \eps}
    =
    \dd A\supbrk{F}\brk{\vec{x}, \eps}
    \cdot
    \vMJ_F\brk{\vec{x}, \eps}
    \> \,,
\end{align}
where the matrix-valued one-form $\dd A(\vec{x}, \eps)$, called \emph{connection matrix}, is given by
\begin{align}
    \dd A\supbrk{F}\brk{\vec{x}, \eps}
    =
    \sum_{a = 1}^7
    A_{x_a}\supbrk{F}\brk{\vec{x}, \eps}
    \>
    \dd x_a
    \> \,.
\end{align}
A key step in computing MIs with this method is to find a basis such that the DEs take a \emph{canonical} form~\cite{Henn:2013pwa}, where certain conditions are met.
The first condition is that $\eps$ factorises from the connection matrix, i.e., that
\begin{equation}
\label{eq:canDE}
\dd A\supbrk{F}\brk{\vec{x}, \eps}  = \eps \, \dd \tilde{A}\supbrk{F}\brk{\vec{x}} \,.
\end{equation}
This makes it straightforward to write the solution in terms of Chen iterated integrals~\cite{Chen:1977oja}.
In the best understood case, we also require that the connection matrix can be written in terms of logarithmic one-forms (alias $\dlog$'s), as
\begin{equation}
\label{eq:dlog}
\dd A\supbrk{F}\brk{\vec{x}, \eps} = \eps \, \sum_i c_i^{(F)} \dlog W_i(\vec{x}) \,,
\end{equation}
where the $c_i^{(F)}$'s are constant rational matrices, and the $W_i(\vec{x})$'s are algebraic functions of the kinematic invariants called \emph{letters}.
The set of all letters is called \emph{alphabet} and captures the singularities of the solution.
The form of the connection matrix in \cref{eq:dlog} gives access to powerful mathematical techniques to handle the solution.

However, we will see that the DEs governing the integral families studied in this work cannot be expressed solely in terms of $\dlog$'s owing to the presence of \emph{elliptic functions}.
While significant progress has been made in constructing canonical DEs beyond the $\dlog$ case~\cite{Adams:2017tga,Adams:2018yfj,Frellesvig:2021hkr,Gorges:2023zgv,Dlapa:2022wdu,Pogel:2022ken,Pogel:2022vat,Frellesvig:2023iwr,Driesse:2024feo,Duhr:2024uid,Becchetti:2025oyb,Duhr:2025lbz,Chen:2025hzq}, the high degree of algebraic complexity of the problem renders this step extremely challenging.
Moreover, we also anticipate that factorising $\eps$ in the connection matrices of the families under study requires the normalisation of the MIs by  \emph{nested square roots}.
Results for MIs of this kind are scarce~\cite{Badger:2024fgb,FebresCordero:2023pww,Becchetti:2025oyb}, and it is in general unclear whether a $\dlog$ form of the connection matrix as in \cref{eq:dlog} can be achieved in such a case.
Finally, even if we did achieve a canonical form, the presence of elliptic periods and nested square roots in the DEs would make an efficient evaluation of their solution significantly more involved and, at the time of writing, unsupported by public tools.
For these reasons, in this work we aim for a less constrained form of the DEs, that will enable the use of generalised power series expansions~\cite{Moriello:2019yhu} with publicly available tools~\cite{Hidding:2020ytt,Armadillo:2022ugh,Prisco:2025wqs} for evaluating the solution.
Furthermore, the resulting integral bases will lead to a significantly more compact representation of the two-loop amplitude, making its computation simpler than one with an arbitrary choice of MIs.

Explicitly, we construct bases of MIs so that the DEs meet the following requirements.
\begin{enumerate}

\item The blocks of the connection matrix coupling MIs of sectors that do not contain nested square roots nor elliptic curves are $\eps$-factorised.
Roughly speaking, the DEs should be as close as possible to being canonical without introducing transcendental functions or nested square roots in the MIs definition.

\item The entries of the connection matrix that are not $\eps$-factorised are polynomial in $\eps$.
We tried to minimise the degree in $\eps$ without introducing elliptic functions.
In practice, the entries of the connection matrices we obtain are at worst quadratic in $\eps$.

\item The connection matrix is free of \emph{spurious denominator factors}.\footnote{In a single case, we will see that allowing for a spurious denominator factor actually yields a better form of the DEs in view of their solution via generalised power series. See \cref{sec:five-point_elliptic}.}
We recall that the common denominator of the entries of the connection matrix defines a hypersurface which contains the singular locus of the solution.
An irreducible polynomial $f(\vec{x})$ in the denominator of the connection matrix is spurious if the solution is finite where $f(\vec{x}) = 0$ for generic boundary values; we talk in this case of an \emph{apparent singularity}.
Having spurious denominator factors in the DEs for the MIs is undesirable as it introduces spurious, unphysical singularities in the amplitudes, and is typically a sign that further simplification can be achieved by a suitable basis transformation.

\item The MIs associated with elliptic curves and nested square roots are non-zero only starting at order $\eps^4$.

\end{enumerate}
The last constraint is important in view of the computation of the two-loop amplitude.
Provided that the latter requires the MIs only up to order $\eps^4$, this feature can be exploited to construct a (potentially over-complete) basis of special functions that allows for the analytic cancellation of ultraviolet and infrared poles, simplifies the expression of the finite remainder, and improves the efficiency of numerical evaluation~\cite{Badger:2024dxo}.
We leave this study to future work, as it also requires information about the amplitude.
Finally, whenever the constraints above are satisfied by multiple choices of MIs, we prefer the one that minimises the algebraic complexity of the connection matrix, as measured by the total polynomial degree in $\vec{x}$ of the numerators of the entries.

In the next subsection we present the general strategy we followed to construct integral bases that satisfy DEs of the form above.
We give more details about the MIs involving nested square roots and elliptic curves in \cref{sec:Nested_square_roots} and \cref{sec:Elliptic_curves}, respectively.
We collect information about the MIs of the other new five-point sectors in \cref{New_Sectors}.

\subsection{General strategy}
\label{sec:general_strategy}

We construct the integral bases by following the strategy outlined in ref.~\cite{Badger:2024fgb}, which we recall here briefly.
Our approach is bottom-up: we first identify a set of MIs satisfying our criteria for sectors with the fewest propagators, and then systematically extend the analysis to sectors with an increasing number of propagators, iteratively refining the DEs until the entire system attains the desired form.
For a given sector $S$, the first step is to select the MIs in $S$ such that the homogeneous connection matrix---i.e., the block of the connection matrix that couples the sector's MIs---takes the form
\begin{equation} \label{eq:hom_guess}
 \dd A_{\operatorname{Hom.}}\supbrk{F,S}\brk{\vec{x}, \eps} = \sum_{a = 1}^7
    \sum_{k = 0}^{k_{\operatorname{max}}}
    \eps^k \> R_{k,a}\supbrk{F,S}\brk{\vec{x}} \> \dd x_a \,,
\end{equation}
where $R_{k,a}\supbrk{F,S}\brk{\vec{x}}$ are matrices of rational functions, and $k_{\operatorname{max}}$ is $2$ for the elliptic sectors and $1$ otherwise. The choice of MIs is guided by known results from the literature and patterns observed in similar processes studied previously~\cite{Gehrmann:2015bfy,Abreu:2020jxa,Badger:2022hno,Badger:2024fgb,FebresCordero:2023pww,Abreu:2024yit}.
For instance, in five-point sectors, a well-suited basis typically includes MIs with $\mu_{ij}$-insertions under the integral sign, cfr.~\cref{eq:muij}.
More details on these choices are provided in the upcoming subsections.
To verify that a chosen basis has a homogeneous connection matrix as in \cref{eq:hom_guess}, we employ finite-field methods~\cite{vonManteuffel:2014ixa,Peraro:2016wsq} to reconstruct only the $\eps$-dependence of the DEs, that is, we fix $\vec{x}$ to random values.
Once this is achieved, we reconstruct the full analytic dependence of the homogeneous connection matrix on $\vec{x}$.
We emphasise that, for the sectors $S$ that do not involve elliptic curves, $R_{0,a}\supbrk{F,S}\brk{\vec{x}}$ is diagonal and has non-zero entries only corresponding
to the MIs that require normalisation by a square root to have $\eps$-factorised homogeneous DEs.
Setting to zero also the diagonal entries and thus achieving the complete factorisation of $\eps$ in the homogeneous block would introduce square roots.
We refrain from doing this at this stage, as this allows us to work with rational expressions for the matrices~$R_{k,a}\supbrk{F,S}\brk{\vec{x}}$.
We boost \soft{FiniteFlow}'s built-in functional reconstruction algorithm with the strategy outlined in refs.~\cite{Abreu:2018zmy,Badger:2021imn}, based on fitting linear relations and determining the denominators from ans\"atze through reconstruction on univariate phase-space slices.

Once the homogeneous blocks of all sectors have the form in \cref{eq:hom_guess},
we reconstruct analytically also the inhomogeneous blocks.
Thanks to our approach to choosing the MIs (illustrated in \cref{sec:Nested_square_roots,sec:Elliptic_curves}), the DEs are at this stage mostly $\eps$-factorised also in the inhomogeneous blocks.
We remove the remaining inhomogeneous entries that are not $\eps$-factorised and do not involve elliptic curves by correcting the relevant MIs by suitable linear combinations of the MIs corresponding to such entries~\cite{Gehrmann:2014bfa}.

Finally, once we have the analytic expression of the connection matrix for the chosen basis, we also introduce the square-root normalisations required to factorise $\eps$ in the diagonal entries corresponding to the MIs that are associated with neither elliptic curves nor nested square roots.
We discuss how we represent the resulting connection matrices in \cref{Results}.

\subsection{Nested square roots}
\label{sec:Nested_square_roots}

\begin{figure}[t]
\centering
\includegraphics[scale=0.4]{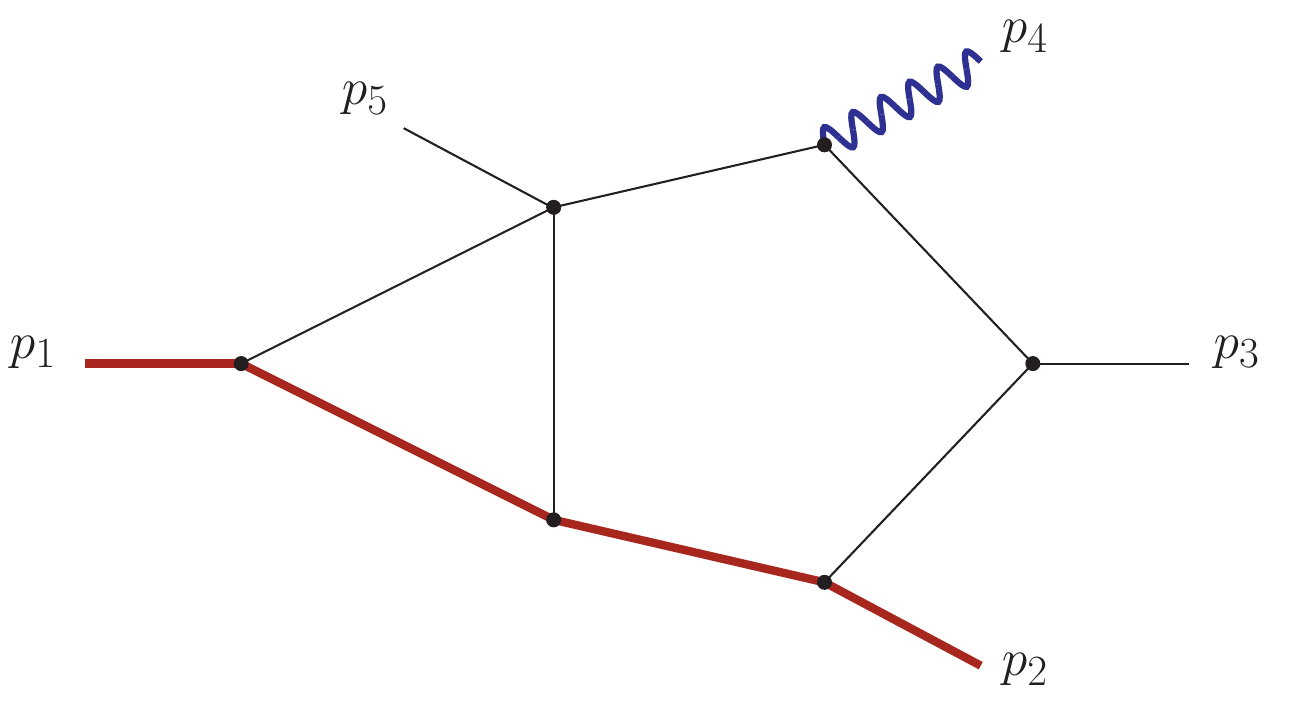}
\caption{Graph representing the sector of family $F_1$ that contains nested square roots. The notation is the same as in \cref{fig:Families}.}
\label{fig:nested}
\end{figure}

In this section, we discuss the sector that contains nested square roots.
It is obtained by pinching the $7$th propagator of family $F_1$, as shown graphically in \cref{fig:nested}, and contains 3~MIs. The appearance of nested square roots as leading singularities was already observed in this `pentagon-triangle' topology in refs.~\cite{Badger:2024fgb,Becchetti:2025oyb}, albeit with different kinematics, as well as in a different topology in ref.~\cite{FebresCordero:2023pww}.

In order to see how nested square roots emerge, we study the loop-by-loop Baikov representation~\cite{Baikov:1996iu, Baikov:1996rk, Frellesvig:2017aai, Frellesvig:2024ymq} of the scalar integral of this sector.
The goal is to construct numerators such that the resulting integrands have simple poles only, and the leading singularities---i.e., the maximal co-dimension residues---are constant~\cite{Cachazo:2008vp,Arkani-Hamed:2010pyv}.
Feynman integrals with these properties are conjectured to satisfy canonical DEs.
We use the package \soft{DlogBasis} to compute the leading singularities~\cite{Wasser:2018qvj,Henn:2020lye}.
By parametrising first the loop which contains fewest propagators (with loop momentum $k_2$) and omitting the overall $\vec{x}$-independent factor we obtain
\begin{equation}
I_{11111101000}^{(F_1)} \propto \int \frac{\dd z_{1}\cdots \dd z_{6} \dd z_{8} \dd z_{9}}{z_{1} \cdots z_{6} z_{8}} \times \frac{{\rm G}(k_2,p_1,k_1)^{-\eps}}{{\rm G}(k_1,p_1)^{1/2-\eps}} \times \frac{{\rm G}(p_1,p_2,p_3,p_4)^{1/2+\eps}}{{\rm G}(k_1,p_1,p_2,p_3,p_4)^{1+\eps}} \,,
\end{equation}
where $z_i$ stems from the $i$th inverse propagator $D_{F_1,i}$.
The integration domain is irrelevant for our purposes.
Taking the maximal cut and setting $\eps=0$ gives
\begin{align}
\label{eq:nested_integrand}
I_{11111101000}^{(F_1)} \bigl|_{\eps=0}^{\text{MC}} \propto  \int \frac{a \, r_1 \, \, \dd z_{9}}{\sqrt{z_{9}-3m_t^2} \sqrt{z_{9} + m_t^2} (2a z_{9}+b+c \, r_1) \bigl(2a z_{9}+b-c \, r_1 \bigr)} \,,
\end{align}
where we recall that $r_1$ is a square root defined in \cref{eq:roots1}, and we introduced the short hands
\begin{align} \label{eq:abc}
\begin{aligned}
a&=(m_t^2 + m_w^2 - s_{23} - s_{34}) (m_t^2 m_w^2 - s_{23} s_{34}) - s_{15} (m_t^2 - s_{23}) (m_w^2 - s_{34}) \,, \\
b&= s_{34} s_{45} \left[ 2 m_t^2 m_w^2 + (m_t^2 - s_{15}) (m_t^2 - s_{23}) + s_{34} (m_t^4 + s_{23}^2 - s_{45}(m_t^2 + s_{23}))  \right] \\
& \ \phantom{=} + s_{34} (s_{12} - 2 m_t^2) \left[ m_w^2 (m_t^2 + s_{23}) + (m_t^2 - s_{23}) (s_{23} - s_{15}) \right] \\
& \ \phantom{=} + 2 m_w^2 (m_t^2 - s_{12}) \left[ m_t^2 m_w^2 + (m_t^2 - s_{15}) (m_t^2 - s_{23})\right]  \, , \\
c & =s_{34}(m_t^2 - s_{23}) \,.
\end{aligned}
\end{align}
We now want to construct numerators for the integrand in \cref{eq:nested_integrand} such that the resulting leading singularities---i.e., the residues of the integrand at all possible simple poles in the integration variable, $z_9$---are constant.
For this purpose, it is useful to note that
\begin{align}
\mu_{11}\bigl|^{\rm MC} = \frac{(2 \, a \, z_9 + b + c \, r_1)(2 \, a \, z_9 + b - c \, r_1)}{2\,a \, r_1^2} \,,
\end{align}
where we recall that $\mu_{11}$ is defined in \cref{eq:muij}.
Then, one can verify that the following integrals have constant leading singularities,
\begin{align}
\label{eq:MC_UT_nested}
\begin{aligned}
\mathrm{J}^{F_1, \, {\rm(MC)}}_{10}&=\eps^4 \, \frac{{\rm NS}_-}{a} \, I_{11111101000}^{(F_1)} \bigl[2a \, z_9+b+c\, r_1 \bigr] \,,  \\
\mathrm{J}^{F_1, \, {\rm(MC)}}_{11}&=\eps^4 \, \frac{{\rm NS}_+}{a} \, I_{11111101000}^{(F_1)} \bigl[2a \, z_9+b - c\, r_1 \bigr] \,,  \\
\mathrm{J}^{F_1,\, \text{(MC)}}_{12}&=\eps^4 \, r_1 \,I_{11111101000}^{(F_1)}\bigl[\mu_{11}\bigr] \,.
\end{aligned}
\end{align}
where the superscript ${\rm(MC)}$ recalls that we are working on the maximal cut, $\mathrm{NS}_{\pm}$ are nested square roots,
\begin{align}
\label{eq:nested_sqrt}
\mathrm{NS}_{\pm} = \pm \frac{\sqrt{b + 6 a m_t^2 \pm c \, r_1} \sqrt{b - 2 a m_t^2 \pm c\, r_1}}{r_1} \, ,
\end{align}
and $a$, $b$ and $c$ are polynomials in $\vec{x}$ defined in \cref{eq:abc}.
We checked that the arguments of the nested square roots are not perfect squares, and that the homogeneous DEs for the basis in \cref{eq:MC_UT_nested} are indeed $\eps$-factorised.
As explained in the introduction to this section, we prefer not to have nested square roots in the basis definition.
We remove them and obtain homogeneous DEs that are linear in $\eps$ by choosing the following basis,
\begin{align}
\begin{aligned}
\mathrm{I}^{F_1, \, \text{(MC)}}_{10} & =\eps^4 \, \frac{s_{45}}{r_1} \, I_{11111101000}^{(F_1)}\bigl[2a z_9 +b \bigr] \,,  \\
\mathrm{I}^{F_1, \, \text{(MC)}}_{11} & =\eps^4 s_{45} \, c \, I_{11111101000}^{(F_1)} \,,\\
\mathrm{I}^{F_1, \, \text{(MC)}}_{12} & =\eps^4 \, r_1 \,I_{11111101000}^{(F_1)}\bigl[\mu_{11}\bigr] \,,
\end{aligned}
\end{align}
where the arbitrary factors of $s_{45}$ in $\text{I}^{F_1, \, \text{(MC)}}_{10} $ and $ \text{I}^{F_1, \, \text{(MC)}}_{11}$ are introduced to give all MIs the same mass dimension.

Reconstructing the DEs beyond the maximal cut shows that contributions from lower sectors have to be included as well in order to factorise $\eps$.
In this analysis, we exclude the block of the connection matrix coupling this sector to the one containing an elliptic curve shown in \cref{fig:ttj_elliptic}; achieving $\eps$-factorisation in this block would require introducing elliptic functions.
In order to catch these extra contributions, we extend the Baikov analysis to sub-maximal cuts.
In particular, leaving $z_{1}$ and $z_{4}$ un-cut is convenient because, on this cut, they appear in the discriminant of $\mu_{11}$---viewed as a polynomial in $z_9$---in an overall perfect square, which factors out of the square root.
The analysis of leading singularities then leads to our final basis for this sector,
\begin{align} \label{eq:basis_nested_I}
\begin{aligned}
\mathrm{I}_{10}^{F_1} & =\eps^4\, \frac{s_{45}}{r_1} \, \left[2aI_{11111101-100}^{(F_1)}+b I_{11111101000}^{(F_1)} + b_1 I_{01111101000}^{(F_1)}+b_2 I_{11101101000}^{(F_1)} \right]  \,, \\
\mathrm{I}_{11}^{F_1} & =\eps^4  \, s_{45} \left[ (s_{23}-m_t^2) \left(s_{34}I_{11111101000}^{(F_1)}-I_{11101101000}^{(F_1)} \right)  - (s_{34}-m_w^2)I_{01111101000}^{(F_1)} \right] \,,  \\
\mathrm{I}_{12}^{F_1} & =\eps^4 \, r_1 \,I_{11111101000}^{(F_1)} \bigl[\mu_{11} \bigr] \,,
\end{aligned}
\end{align}
where $b_1$ and $b_2$ are 3rd-degree polynomials of $\vec{x}$,
\begin{align}
\begin{aligned}
b_1&=s_{23} \left[m_w^2 ( s_{12} + s_{34}) + s_{34} (s_{12} - s_{34} - 2 s_{45}) \right] + m_t^2 m_w^2 (s_{45} - s_{34} - 2 s_{12})\\
& \ \phantom{=} + m_t^2 s_{34} (s_{34} + s_{45}) - (m_w^2 - s_{34}) \left[s_{12} (m_w^2 - s_{15}) + (s_{15} - s_{34}) s_{45} \right] \,,
\\
b_2&= s_{12} \left[m_t^2 (m_w^2 - s_{15} + s_{23}) + s_{23} (m_w^2 + s_{15} - s_{23} - 2 s_{34}) \right] - m_t^4 (s_{34} + s_{45})\\
&\ \phantom{=} + m_t^2 (s_{15} + s_{23} + s_{34} - 2 m_w^2) s_{45} - s_{23} \left[s_{23} s_{34} + (s_{15} - s_{34}) s_{45} \right]  + 2 m_t^2 s_{23} s_{34} \,.
\end{aligned}
\end{align}
We verified by numerical evaluation with \soft{AMFlow}~\cite{Liu:2022chg} that $\mathrm{I}_{12}^{F_1}$ vanishes up to (and including) order $\eps^4$, while $\mathrm{I}_{10}^{F_1}$ and $\mathrm{I}_{11}^{F_1}$ are non-zero starting from $\eps^4$,
The factorisation of $\eps$ is achieved (except for the entries that couple these MIs to the elliptic sector in \cref{fig:ttj_elliptic}, which are linear in $\eps$) by the following transformation,
\begin{equation}
\label{eq:UT_rotation_nested}
\begin{pmatrix} \mathrm{J}_{10}^{F_1} \\ \mathrm{J}_{11}^{F_1} \\ \mathrm{J}_{12}^{F_1} \end{pmatrix} = \begin{pmatrix} \frac{r_1 {\rm NS}_-}{a\,s_{45}} & 0 & 0\\ 0 & \frac{r_1 {\rm NS}_+}{a\,s_{45}} & 0 \\ 0 & 0 & 1 \end{pmatrix} \cdot \begin{pmatrix} 1 & -1 & 0\\  1 & 1 & 0 \\ 0 & 0 & 1 \end{pmatrix} \cdot \begin{pmatrix} {\rm I}_{10}^{F_1} \\ {\rm I}_{11}^{F_1} \\ {\rm I}_{12}^{F_1} \end{pmatrix} \,.
\end{equation}

We conclude this section with some remarks concerning nested square roots.
First, as expected, ${\rm NS}_{\pm}$ become simple square roots ($r_1$ times a rational function of $\vec{x}$) in the limit $m_t^2 \to 0$.
Second, it is useful to characterise the singular locus of ${\rm NS}_{\pm}$ as an algebraic variety, i.e., as the vanishing locus of a polynomial.
The associated polynomial is given by the product $r_1^4 {\rm NS}_{+}^2 {\rm NS}_{-}^2$, which factors into a number of irreducible polynomials with degree ranging from $2$ to $5$.
The latter appear in the denominators of the entries of the connection matrix.
Finally, it has been known for a long time that MIs can be chosen so that they have a well-defined parity, either even or odd, with respect to flipping the sign of each square root.
For example, from \cref{eq:basis_nested_I} we can see that ${\rm I}_{10}^{F_1}$ and ${\rm I}_{12}^{F_1}$ are odd w.r.t.\ $r_1$, whereas ${\rm I}_{11}^{F_1}$ is even.
This picture becomes more complicated in the presence of nested square roots.
Indeed, the transformation required to factorise $\eps$, in \cref{eq:UT_rotation_nested}, mixes integrals with different $r_1$-parity.
In ref.~\cite{Becchetti:2025oyb}, the authors proposed that the natural generalisation of the even/odd parity to nested square roots is to arrange all objects into `duplets' of elements that transform into each other upon flipping the sign of the interior root, i.e., a 2-dimensional representation of $\mathbb{Z}_2$.
We find that this is possible in our case as well: while ${\rm J}^{F_1}_{10}$ and ${\rm J}^{F_1}_{11}$ are odd w.r.t.\ to the sign of the exterior root of ${\rm NS}_-$ and ${\rm NS}_+$, respectively, they form a duplet w.r.t.\ the sign of the interior root, $r_1$, as do ${\rm NS}_-$ and ${\rm NS}_+$.

\subsection{Elliptic curves}
\label{sec:Elliptic_curves}

\begin{figure}[t]
\centering
\begin{subfigure}[b]{0.8\textwidth}
\centering
\includegraphics[scale=0.4]{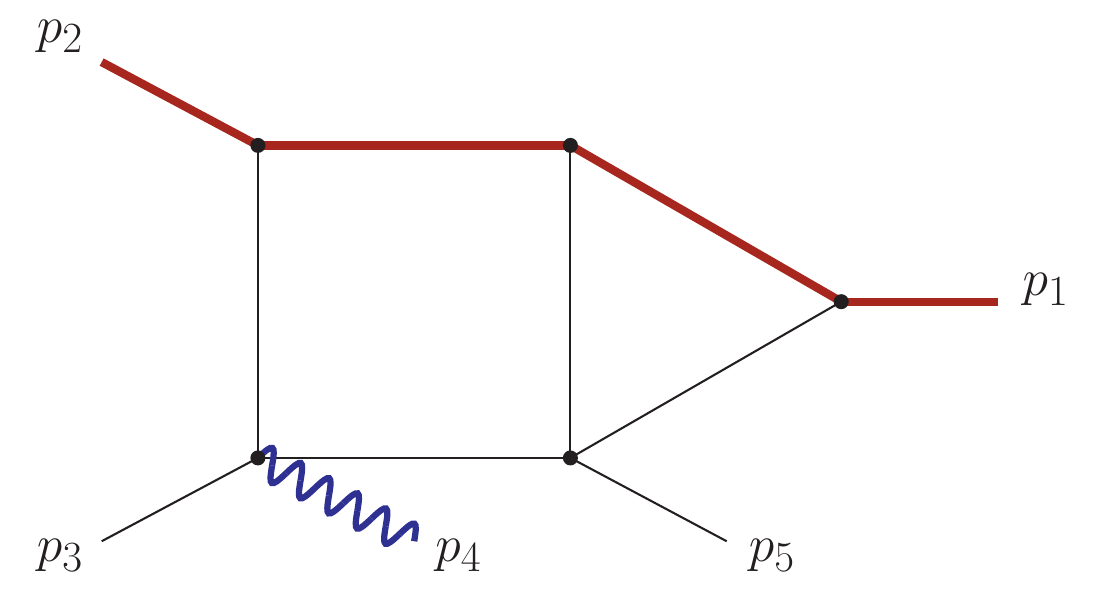}
\caption{Four-point elliptic sector of $F_1$ already studied in refs.~\cite{Badger:2024fgb,Becchetti:2025oyb}.}
\label{fig:ttj_elliptic}
\end{subfigure}
\hfill
\begin{subfigure}[b]{0.496\textwidth}
\centering
\includegraphics[scale=0.4]{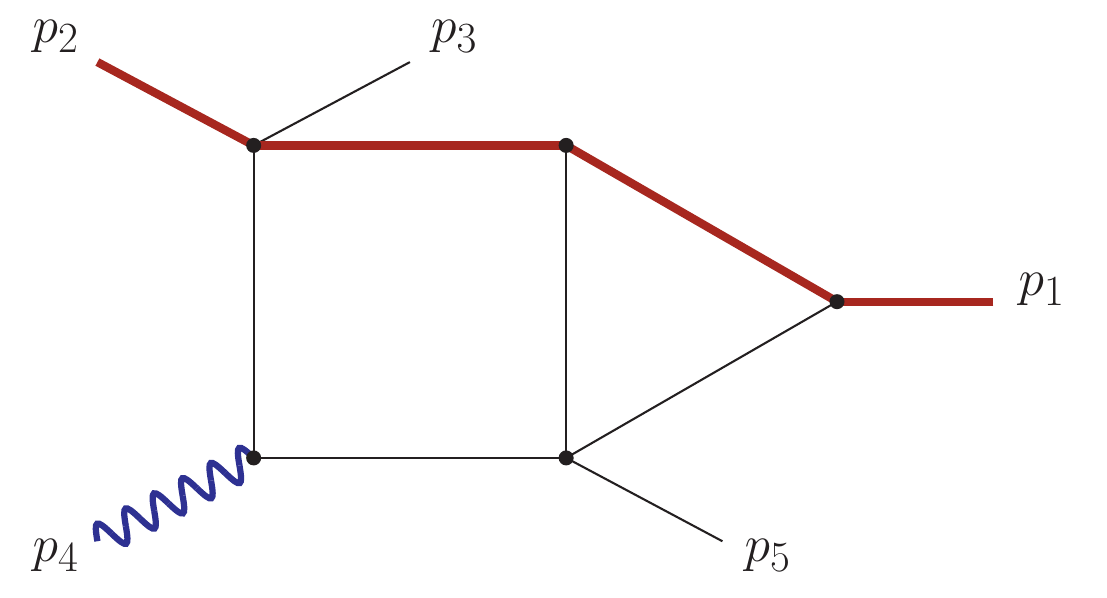}
\caption{Four-point elliptic sector of $F_1$.}
\label{fig:4pt_elliptic}
\end{subfigure}
\hfill
\begin{subfigure}[b]{0.496\textwidth}
\centering
\includegraphics[scale=0.4]{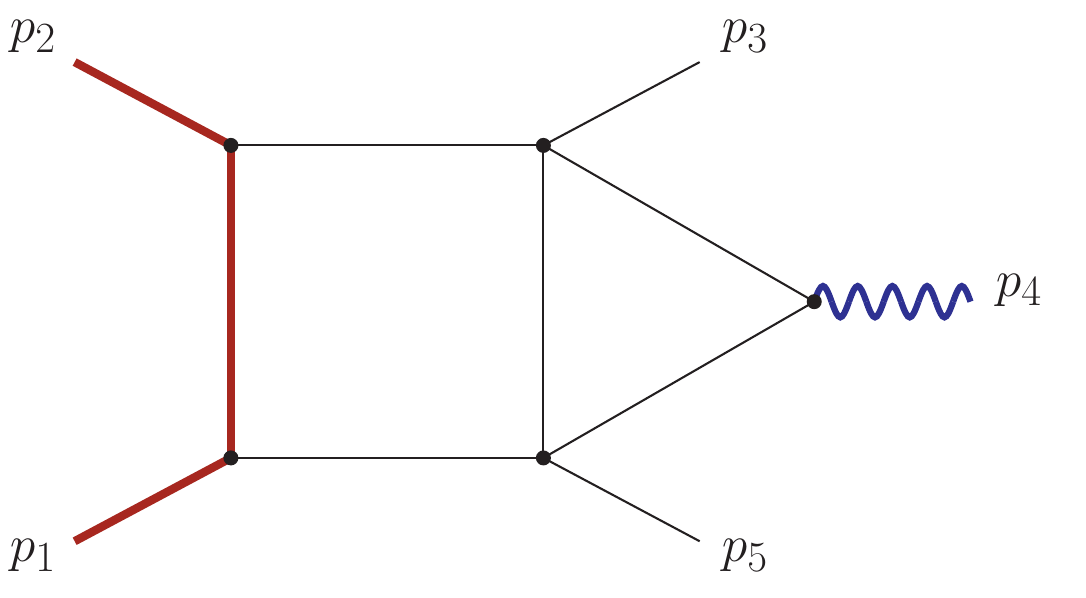}
\caption{Five-point elliptic sector of $F_2$ and $F_3$.}
\label{fig:5pt_elliptic}
\end{subfigure}
\caption{Feynman graphs of the sectors containing elliptic curves. The notation is the same as in \cref{fig:Families}.}
\label{fig:elliptics}
\end{figure}

This section is devoted to the three sectors that contain elliptic curves: we discuss how we identified the underlying elliptic curves, and how we chose appropriate MIs.
One elliptic sector, appearing in $F_1$ (MIs $\mathrm{I}_{i}^{F_1}$ with $i=30,31,32$) and shown in \cref{fig:ttj_elliptic}, has four-point kinematics and is identical to the elliptic sector studied in refs.~\cite{Badger:2024fgb,Becchetti:2025oyb}.
Although the DEs for this sector could be brought to $\eps$-factorised form~\cite{Becchetti:2025oyb}, we refrain from doing so here for the reasons outlined in the introduction to \cref{sec:Bases_and_DEs}.
We adopt the same basis proposed in ref.~\cite{Badger:2024fgb}, and do not discuss this sector further.
The second elliptic sector, shown in \cref{fig:4pt_elliptic}, is similar to the first one: it also has four-point kinematics, appears only in $F_1$ and has $3$ MIs ($\mathrm{I}_{i}^{F_1}$, with $i=33,34,35$); we discuss it in \cref{sec:four-point_elliptic}.
The third elliptic sector, shown in \cref{fig:5pt_elliptic}, is considerably more complicated: it has five-point kinematics, appears in both $F_2$ (MIs $\mathrm{I}_{i}^{F_2}$, with $i=17,\ldots,23$) and $F_3$ (MIs $\mathrm{I}_{i}^{F_3}$, with $i=20,\ldots,26$), and has $7$ MIs; we discuss it in \cref{sec:five-point_elliptic}.
We emphasise that the three elliptic curves associated with these sectors are not isomorphic to each other: they depend on different sets of kinematic invariants and we verified that they have different $j$-invariants~\cite{Lang1987-zb}
(see \cref{app:elliptic_details} for an explicit computation of a $j$-invariant).

\subsubsection{Four-point elliptic sector}
\label{sec:four-point_elliptic}

In this subsection we study the sector of $F_1$ defined by the Feynman graph in \cref{fig:4pt_elliptic}.
It contains $3$ MIs and has four-point kinematics; explicitly, it depends on the subset $\bigl\{s_{23},s_{45},s_{15},m_t^2,m_w^2\bigr\}$ of the variables $\vec{x}$.
Since this sector is similar to the one studied in ref.~\cite{Badger:2024fgb} (see \cref{fig:ttj_elliptic}), we followed the same approach to choose the following MIs:
\begin{align}
\begin{aligned}
\mathrm{I}_{33}^{F_1}&=\eps^4 (1-2\eps) (s_{15} -
   m_t^2) \left(I_{10111101001}^{(F_1)} - I_{10110101000}^{(F_1)} - s_{23} I_{10111101000}^{(F_1)} \right) \,, \\
\mathrm{I}_{34}^{F_1}&=
\eps^4 \, r_6 \left(I_{10111101000}^{(F_1)} -I_{00111101000}^{(F_1)} - m_t^2 I_{10111101000}^{(F_1)} \right) \,, \\
\mathrm{I}_{35}^{F_1}&= \eps^4 s_{45} (s_{15} - m_t^2) \, I_{10111101000}^{(F_1)} \,.
\end{aligned}
\end{align}
As in ref.~\cite{Badger:2024fgb}, the kinematic normalisation factors have been chosen heuristically so that all MIs have the same mass dimension.
This choice of basis ensures that the connection matrix has the desired form.
In particular, we have the following $\eps$ structure on the maximal cut,
\begin{align}
\dd \begin{pmatrix} {\rm I}_{33}^{F_1} \\ {\rm I}_{34}^{F_1} \\ {\rm I}_{35}^{F_1}  \end{pmatrix} = \begin{pmatrix} \ast +  \ast \eps  & \ast \eps (2\eps - 1) & (2\eps - 1) (\ast+\ast \eps) \\ \ast & \ast \eps & \ast + \ast \eps \\  \ast & \ast \eps & \ast + \ast \eps \end{pmatrix} \cdot \begin{pmatrix} {\rm I}_{33}^{F_1} \\ {\rm I}_{34}^{F_1} \\ {\rm I}_{35}^{F_1}  \end{pmatrix} + \cdots \,,
\end{align}
where the asterisks denote distinct non-zero terms, and the ellipsis stands for the omitted sub-sector contributions.
The sub-sector entries of the connection matrix are $\eps$-factorised for the differentials of $\mathrm{I}_{34}^{F_1}$ and $\text{I}_{35}^{F_1}$, and depend quadratically on $\eps$ for the differential of $\text{I}_{33}^{F_1}$.
Furthermore, the chosen MIs of this sector are non-zero only starting from order $\eps^4$, as we verified by evaluating them numerically in several random phase-space points with \soft{AMFlow}~\cite{Liu:2022chg}.
The similarity with the elliptic sector of refs.~\cite{Badger:2024fgb,Becchetti:2025oyb} suggests that this choice of basis is also a suitable starting point to apply the strategy of refs.~\cite{Gorges:2023zgv} in order to achieve the complete factorisation of $\eps$, as done in ref.~\cite{Becchetti:2025oyb}.

In order to reveal the elliptic curve underlying this sector, we analyse the loop-by-loop Baikov representation of its scalar integral ($\mathrm{I}_{35}^{F_1}$, up to an overall factor) as in \cref{sec:Nested_square_roots}.
By starting from the $k_2$ loop, omitting the overall kinematics-independent factor, taking the maximal cut and setting $\eps=0$ we obtain
\begin{equation}
\label{eq:MC_belli}
I_{10111101000}^{(F_1)} \bigl|^{\rm MC}_{\eps=0} \propto \int \frac{\dd z}{\sqrt{\mathcal{P}_{4\text{-pt.}}(z)}} \,,
\end{equation}
where we omitted the integration domain, the integration variable $z$ is the $9$th inverse propagator $D_{F_1,9}$, and
$\mathcal{P}_{4\text{-pt.}}(z)$ is a quartic polynomial in $z$,
\begin{align} \label{eq:P4pt}
 \mathcal{P}_{4\text{-pt.}}(z) = (z + m_t^2)(z - 3 m_t^2)\mathcal{P}_2(z) \,,
\end{align}
with
\begin{align}
\begin{aligned}
\mathcal{P}_2(z)&=\left(z - 3 m_t^2 \right) \left(z + m_t^2 \right) m_w^4 + \left[\left(z + m_t^2\right) (s_{15} - s_{23}) + \left(m_t^2 - s_{15}\right) s_{45} \right]^2 \\
& \ \phantom{=} + 2 \left(z + m_t^2 \right) m_w^2 \left[ s_{15} \left(m_t^2 - 2 s_{23} + s_{45} - z \right) + m_t^2 \left(s_{23} + s_{45} - 2 m_t^2 \right) - z s_{23}\right] \,.
\end{aligned}
\end{align}
The quartic polynomial $\mathcal{P}_{4\text{-pt.}}(z)$ defines an elliptic curve in the $(z,y)$ plane, as
\begin{align}
\label{eq:baby_curve}
E_{4\text{-pt.}} : \qquad y^2 = \mathcal{P}_{4\text{-pt.}}(z) \,.
\end{align}
Then, depending on the integration contour, the maximal cut in \cref{eq:MC_belli} can be expressed as a linear combination of the two periods of the elliptic curve, given by complete elliptic integrals of the first kind.
It is interesting to look at the discriminant of $\mathcal{P}_{4\text{-pt.}}(z)$, which characterises the geometric properties and singularities of the curve: if the discriminant vanishes, the roots degenerate and the geometry underlying the integral reduces to a genus-zero surface.
Up to the overall constant, it is given by
\begin{align}
\begin{aligned}
& \Delta_{E_{4\text{-pt.}}} = m_t^4 \, m_w^2 \, s_{45}^4 \, \left(s_{15}-m_t^2 \right)^4  \left[s_{15} s_{23} + m_t^2 \left(m_w^2 - s_{15} - s_{23}\right) + m_t^4\right] \, {\rm G}(p_1, p_{23}, p_4)  \\
& \phantom{\Delta_E} \times \left\{ 8 m_t^2 m_w^2 \left(m_t^2 + s_{15} \right) \left[2 \left(m_t^2 + s_{23}\right) - s_{45}\right] - \left[s_{15} s_{45} - m_t^2 (4 s_{15} - 4 s_{23} + s_{45})\right]^2\right\}^2 \,,
\end{aligned}
\end{align}
where ${\rm G}(p_1, p_{23}, p_4) $ is an irreducible degree-3 polynomial in $\vec{x}$.
We therefore see that the roots are distinct for generic $\vec{x}$, and that---as expected---they degenerate in both massless limits, $m_t^2 \to 0$ and $m_w^2 \to 0$.
All irreducible factors in $\Delta_{E_{4\text{-pt.}}}$ define potential singularities of the integral, and indeed appear in the denominators of the entries of the connection matrix.

\subsubsection{Five-point elliptic sector}
\label{sec:five-point_elliptic}

We now turn to the five-point elliptic sector defined by the Feynman graph in \cref{fig:5pt_elliptic}.
The integrals of this sector depend on all $7$ variables $\vec{x}$; this is the first time that an elliptic sector depending on so many kinematic invariants has been studied, and we will see that the increase in complexity is considerable.
The sector consists of $7$ MIs, and appears in both $F_2$ and $F_3$.
We focus the discussion on the MIs for $F_2$ ($\text{I}^{F_2}_{i}$ with $i=17,\ldots,23$); we choose the MIs for $F_3$ analogously ($\text{I}^{F_3}_{i}$ with $i=20,\ldots,26$).

Our choice for the MIs of this sector is the following:
\begin{align}
\label{eq:elliptic_sector_MIs}
\begin{aligned}
\mathrm{I}^{F_2}_{17} &= \eps^4 \frac{r_1^2}{s_{12}} \frac{\partial}{\partial s_{15}} I^{(F_2)}_{11100111000} \,,  \\
\mathrm{I}^{F_2}_{18} &= \eps^4 s_{12} \left(s_{12} - 4 m_t^2\right) I^{(F_2)}_{11100111000} \,,  \\
\mathrm{I}^{F_2}_{19} &= \eps^3 m_w^2 r_9  I^{(F_2)}_{11100211100} \,,  \\
\mathrm{I}^{F_2}_{20} &= \eps^3 m_w^2 s_{12} \left(m_t^2 - s_{15}\right) I^{(F_2)}_{11100211000} \,, \\
\mathrm{I}^{F_2}_{21} &= \eps^3 m_w^2 s_{12} \left(m_t^2 - s_{23}\right) I^{(F_2)}_{11100121000} \,,  \\
\mathrm{I}^{F_2}_{22} &= \eps^3 r_1 I^{(F_2)}_{11100112000}\bigl[\mu_{12}\bigr] \,,  \\
\mathrm{I}^{F_2}_{23} &= \eps^3 r_1 I^{(F_2)}_{11100112000}\bigl[\mu_{11}\bigr] \,.
\end{aligned}
\end{align}
We denote them cumulatively by $\textbf{I}^{F_2}_{17-23} = \bigl(\mathrm{I}^{F_2}_{17}, \ldots, \mathrm{I}^{F_2}_{23} \bigr)^{\top}$.
With this basis, the DEs take the desired form.
In particular, on the maximal cut they have the following $\eps$-dependence:
\begin{align} \label{eq:maxcut_DE_5pt_elliptic}
\dd \, \textbf{I}^{F_2}_{17-23} =
\begin{pmatrix}
\ast +  \eps  \ast & \ast + \ast \eps + \ast \eps^2 & \ast \eps + \ast \eps^2 & \ast \eps + \ast \eps^2 & \ast \eps  + \ast \eps^2 & \ast \eps + \ast \eps^2 & \ast \eps + \ast \eps^2 \\
\ast & \ast+ \ast \eps & \ast \eps  & \ast \eps & \ast \eps & \ast \eps & \ast \eps \\
\ast & \ast+ \ast \eps & \ast \eps  & \ast \eps & \ast \eps & \ast \eps & \ast \eps \\
\ast & \ast+ \ast \eps & \ast \eps  & \ast \eps & \ast \eps & \ast \eps & \ast \eps \\
\ast & \ast+ \ast \eps & \ast \eps  & \ast \eps & \ast \eps & \ast \eps & \ast \eps \\
\ast & \ast+ \ast \eps & \ast \eps  & \ast \eps & \ast \eps & \ast \eps & \ast \eps \\
0 & \ast \eps & \ast \eps  & \ast \eps & \ast \eps & \ast \eps & \ast \eps
\end{pmatrix} \cdot \textbf{I}^{F_2}_{17-23} \,,
\end{align}
where we omit the sub-sector contributions, and the asterisks denote distinct non-zero terms.
From \cref{eq:maxcut_DE_5pt_elliptic} we see that only $\mathrm{I}^{F_2}_{17}$ and $\mathrm{I}^{F_2}_{18}$ are coupled at $\eps=0$ on the maximal cut.
Beyond the maximal cut, the entries of the connection matrix are $\eps$-factorised for the differential of all MIs of this sector except for $\mathrm{I}^{F_2}_{17}$, for which they are quadratic in $\eps$.
We chose $\mathrm{I}^{F_2}_{22}$ and $\mathrm{I}^{F_2}_{23}$ in analogy with the MIs of a similar sector appearing in the two-loop five-point integral families with one off-shell leg and massless internal propagators~\cite{Abreu:2020jxa}.
We constructed $\mathrm{I}^{F_2}_{19}$, $\mathrm{I}^{F_2}_{20}$ and $\mathrm{I}^{F_2}_{21}$ by analysing the leading singularities in the loop-by-loop Baikov representation, similarly to the discussion in \cref{sec:Nested_square_roots,sec:four-point_elliptic}.
Justifying our choice for $\mathrm{I}^{F_2}_{17}$ and $\mathrm{I}^{F_2}_{18}$ instead requires a deeper analysis of the Baikov parametrisation, that also reveals the underlying elliptic curve.

As in \cref{sec:Nested_square_roots,sec:four-point_elliptic}, we analyse the loop-by-loop Baikov parametrisation of the scalar integral of this sector ($\mathrm{I}^{F_2}_{18}$, up to an overall factor).
We take the maximal cut, neglect the overall kinematics-independent factor, and set $\eps=0$, obtaining
\begin{equation}
I^{(F_2)}_{11100111000} \bigl|^{\rm MC}_{\eps=0} \propto \int \dd z_4 \wedge \dd z_9 \, \mathcal{J}(z_4, z_9) \,,
\end{equation}
where $\mathcal{J}(z_4, z_9)$ is an algebraic function.
In contrast to \cref{sec:Nested_square_roots,sec:four-point_elliptic}, we see that two integration variables are left.
Testing whether the integrand has only simple poles near every singular point thus requires a more sophisticated analysis.
One way to do this is to rewrite the integrand as a linear combination of $\dlog$ forms~\cite{Arkani-Hamed:2010pyv,Wasser:2018qvj,Henn:2020lye}, which make the simple-pole property manifest globally.
Integrating $\mathcal{J}(z_4, z_9)$ in $z_9$ gives a linear combination of logarithms;
we can thus turn $\dd z_9$ into a linear combination of $\dlog$'s, as
\begin{equation}
\label{eq:Baikov_fam2_sec7_2}
I^{(F_2)}_{11100111000} \bigl|^{\rm MC}_{\eps=0} \propto  \int \left( C(z_4) \, \dd z_4 \wedge \dlog\left[\alpha(z_4, z_9) \right] - C(z_4)^{\dagger} \, \dd z_4 \wedge \dlog [ \alpha(z_4, z_9)^{\dagger} ] \right) \, ,
\end{equation}
where we have introduced the following short-hand notation for the $r_1$-sign flip,
\begin{equation}
f^{\dagger} \equiv \left. f \right|_{r_1 \to - r_1} \,.
\end{equation}
Note that the integrand in \cref{eq:Baikov_fam2_sec7_2} is manifestly odd w.r.t.\ the $r_1$-sign flip, but the integral is not.
The explicit expression of the algebraic function $\alpha(z_4,z_9)$ in \cref{eq:Baikov_fam2_sec7_2} is irrelevant, but $C(z_4)$ contains a nested square root.
In order to proceed, we change integration variable from $z_4$ to $z$ so as to rationalise the interior root.
We achieve this by the transformation
\begin{equation}
z_4 \mapsto \frac{4 m_t^2 s_{45} - 2 s_{12}(m_t^2 + s_{23}) + 2 s_{12} (m_t^2 - s_{23})^2 z}{4 m_t^2 - s_{12} + s_{12} (m_t^2 - s_{23})^2 z^2} \> ,
\end{equation}
which we obtained with the \soft{Mathematica} package \soft{RationalizeRoots}~\cite{Besier:2019kco}.
We obtain
\begin{equation}
\label{eq:elliptic_monster}
I^{(F_2)}_{11100111000} \bigl|^{\rm MC}_{\eps=0} \propto (s_{23}-m_t^2){\rm G}(p_1,p_2,p_3) \int \left( \frac{\dd z  \wedge \dlog [\alpha'(z, z_9)]}{\sqrt{\mathcal{P}_{5\text{-pt.}}(z)}} - \frac{\dd z \wedge \dlog [\alpha'(z, z_9)^{\dagger}]}{\sqrt{\mathcal{P}_{5\text{-pt.}}(z)^{\dagger}}}  \right) \,,
\end{equation}
where $\mathcal{P}_{5\text{-pt.}}(z)$ is a quartic polynomial in $z$ with non-vanishing discriminant.
At first sight, the integrand in \cref{eq:elliptic_monster} might appear to contain two elliptic curves,
\begin{align}
\label{eq:5pt_ell_curves}
E_{5\text{-pt.}} : \ y^2 = \mathcal{P}_{5\text{-pt.}}(z) \,, \qquad \quad
E_{5\text{-pt.}}^{\dagger} : \ y^2 = \mathcal{P}_{5\text{-pt.}}(z)^{\dagger} \,,
\end{align}
related by flipping the sign of $r_1$.
Computing their $j$-invariants reveals that they are isomorphic to each other, and thus that they do not depend on the sign of $r_1$.
Therefore, the two elliptic curves can be parametrised in terms of the same Weierstra\ss{} equation~\cite{Lang1987-zb},
\begin{align}
E_{5\text{-pt.}} \sim E_{5\text{-pt.}}^{\dagger}: \ y^2 = \mathcal{P}^{\text{W.}}_{5\text{-pt.}}(z) \,,
    \label{eq:elliptic_monster_curves}
\end{align}
with
\begin{align} \label{eq:P5ptW}
\mathcal{P}^{\text{W.}}_{5\text{-pt.}}(z)  = 4 \, z^3 + g_1 \, z + g_0 \,,
\end{align}
where the coefficients $g_1$ and $g_0$ are polynomials in $\vec{x}$, with irreducible factors of degree up to $9$ and $6$, respectively, and are free of $r_1$.
This can be achieved via an implicit M\"obius transformation $z \mapsto \brk{a_1 + a_2 z} / \brk{a_3 + a_4 z}$, where $a_i$ are functions of the roots of either $\mathcal{P}_{5\text{-pt.}}(z)$ or $\mathcal{P}_{5\text{-pt.}}(z)^{\dagger}$.
As a result, the maximal cut in \cref{eq:elliptic_monster} takes the form
\begin{align}
    \label{eq:elliptic_monster_weierstrass}
    I^{(F_2)}_{11100111000} \bigl|^{\rm MC}_{\eps=0} \propto  (s_{23}-m_t^2) \, {\rm G}(p_1,p_2,p_3) \int {
        \frac{\dd z}{\sqrt{\mathcal{P}^{\text{W.}}_{5\text{-pt.}}(z)}} \wedge \dlog\sbrk{\alpha''(z, z_9)}
    } \,,
\end{align}
which makes it clear that, up to the normalisation, $I^{(F_2)}_{11100111000}$ (and thus $\mathrm{I}^{F_2}_{18}$) is a good MI for this sector.
Furthermore, as expected, the roots of $\mathcal{P}^{\text{W.}}_{5\text{-pt.}}(z)$ degenerate and the integrand of $\mathrm{I}^{F_2}_{18}$ can be cast into a fully $\dlog$ form with the leading singularities known in the literature~\cite{Abreu:2020jxa, Badger:2024fgb} in both massless limits, $m_w^2 \to 0$ and $m_t^2 \to 0$.
We emphasise that all these manipulations are made challenging by the complexity of the expression of $\mathcal{P}_{5\text{-pt.}}(z)$;
just the coefficient of $z^4$, for example, counts $425$ terms once expanded.
Moreover, the discriminant of $\mathcal{P}_{5\text{-pt.}}(z)$ contains an irreducible degree-14 polynomial in $\vec{x}$ with $2547$ monomials, which appears in the denominators of the entries of the connection matrix.
Given the somewhat surprising complexity of this component of the singular locus of $I^{(F_2)}_{11100111000}$, we additionally verified with \soft{SOFIA}~\cite{Correia:2025yao} that it can also be derived from the Landau equations~\cite{Bjorken:1959fd,Landau:1959fi,10.1143/PTP.22.128,Panzer:2014caa,Fevola:2023kaw,Fevola:2023fzn,Caron-Huot:2024brh}.
We give more details about these computations in \cref{app:elliptic_details}, and provide the explicit expression of the Weierstra\ss{} cubic $\mathcal{P}^{\text{W.}}_{5\text{-pt.}}(z)$ in the ancillary files~\cite{zenodo}.

Finally, we have to justify the choice of $\mathrm{I}^{F_2}_{17}$ in \cref{eq:elliptic_sector_MIs}.
We verified via numerical evaluation with \soft{AMFlow} that $\mathrm{I}^{F_2}_{18}$ is non-zero starting from order $\eps^4$.
Choosing $\mathrm{I}^{F_2}_{17}$ as a derivative of $\mathrm{I}^{F_2}_{18}$ thus guarantees that it has the latter property as well.
The variable in the derivative is chosen by trial so as to achieve the most compact expression of the connection matrix.
However, this choice leads to the appearance of a spurious degree-$9$ denominator factor in the connection matrix.
This feature is undesirable and can be removed (without introducing other spurious denominator factors) by replacing $\mathrm{I}^{F_2}_{11}$ with
\begin{equation}
\tilde{\rm I}^{F_2}_{17}=\eps^2 (1 - 2 \eps) (1 - 3 \eps) s_{12} \, I^{(F_2)}_{111-10112000}\bigl[\mu_{11}\bigr] \,.
\end{equation}
However, this basis change has two negative effects.
First, $\tilde{\rm I}^{F_2}_{17}$ is non-zero already at order $\eps^2$.
Second, the DEs would become less compact, more coupled at $\eps=0$, and would contain more non-$\dlog$ one-forms and fewer $\dlog$ one-forms (see \cref{Results}).
For these reasons, we prefer to adopt the basis given in \cref{eq:elliptic_sector_MIs}.

\section{Differential equations}
\label{Results}

In this section, we detail the construction and solution of the DEs governing the integral bases defined
in~\cref{sec:Bases_and_DEs}.
We first discuss how we represent the connection matrices in terms of sets of linearly independent one-forms, separating the logarithmic ones and highlighting the associated algebraic structures.
We then discuss the numerical solution of the DEs using the method of generalised series expansions.

\subsection{Representation of the connection matrices}
\label{sec:DEs_Form}

In this section we discuss how we cast the connection matrices $\dd A_{ij}\supbrk{F}\brk{\vec{x}, \eps}$ computed analytically in the previous section into a form that highlights analytic properties and minimises repeated patterns.
Explicitly, the target form is the following:
\begin{align}
    \dd A\supbrk{F}\brk{\vec{x}, \eps}
    =
    \sum_{k = 0}^2
    \>
    \eps^k \Bigsbrk{
        \sum_{\alpha}
        c_{k \alpha}\supbrk{F} \> \dlog\bigbrk{W_\alpha\brk{\vec{x}}}
        +
        \sum_{\beta}
        d_{k \beta}\supbrk{F} \> \omega_{\beta}\brk{\vec{x}}
    }
    \>.
    \label{eq:DE_dlog_one_form}
\end{align}
Here, the matrices $c_{k \alpha}\supbrk{F}$ and $d_{k \beta}\supbrk{F}$ have rational, constant entries, the $W_\alpha$'s are algebraic functions of the kinematic variables $\vec{x}$ (called letters),
while $\{\omega_{\beta}\brk{\vec{x}}\}$ is a set of $\mathbb{Q}$-linearly independent one-forms that spans the connection matrix elements which cannot be expressed as $\dlog$'s.
We emphasise that most of the `non-$\dlog$' one-forms $\omega_{\beta}\brk{\vec{x}}$ are not closed, and hence there exist no functions of which they are the exterior derivative (see below for more details on the ones that are instead closed).
All one-forms, $\dlog$ and not, are graded according to their even/odd parity with respect to each square root.

In order to reach the form in \cref{eq:DE_dlog_one_form}, we begin by explicitly gathering the differentials for each entry
of the connection matrices in \cref{eq:DE_one_form}, making the
$\eps$-dependence and square-root content manifest by arranging them as
\begin{align}
    \dd A_{ij}\supbrk{F}\brk{\vec{x}, \eps}
    =
    \cC_{ij}\supbrk{F}\brk{\vec{x}}
    \>
    \sum_{a = 1}^7
    \sum_{k = 0}^2
    \eps^k \> f_{ij \, ka}\supbrk{F}\brk{\vec{x}} \> \dd x_a
    \label{eq:DE_charges_eps}
    \>,
\end{align}
where the tensor components $f_{ij \, ka}\supbrk{F}$ are rational functions $\vec{x}$, while $\cC_{ij}\supbrk{F}$ is the product of the square roots associated with the $i$th and $j$th MIs;
if both MIs are even w.r.t.\ all square roots, $\cC_{ij}\supbrk{F} = 1$.
We refer to the $\cC_{ij}\supbrk{F}$'s as (square root) \emph{charges}: the connection matrix element $\dd A_{ij}\supbrk{F}$ is odd w.r.t.\ flipping the sign of $\cC_{ij}\supbrk{F}$;
connection-matrix elements with trivial charge, $\cC_{ij}\supbrk{F}=1$, are rational.
In addition to the $13$ individual square roots listed in \cref{eq:roots1,eq:roots2},\footnote{We recall that $r_3$ does not appear in the families considered here; see \cref{Kinematics_and_Families}.} we find the following set of 33 quadratic charges across the three families:
\begin{align}
\begin{aligned}
    \Bigbrc{
        &
        r_1 \, r_2, \>
        r_1 \, r_4, \>
        r_1 \, r_5, \>
        r_4 \, r_5, \>
        r_1 \, r_6, \>
        r_5 \, r_6, \>
        r_1 \, r_7, \>
        r_2 \, r_7, \>
        r_6 \, r_7, \>
        r_1 \, r_8, \>
        r_1 \, r_9, \>
        \\
        &
        r_2 \, r_9, \>
        r_4 \, r_9, \>
        r_5 \, r_9, \>
        r_7 \, r_9, \>
        r_8 \, r_9, \>
        r_1 \, r_{10}, \>
        r_6 \, r_{10}, \>
        r_7 \, r_{10}, \>
        r_9 \, r_{10}, \>
        r_1 \, r_{11}, \>
        r_5 \, r_{11}, \>
        \\
        &r_6 \, r_{11}, \>
        r_9 \, r_{11}, \>
        r_1 \, r_{12}, \>
        r_7 \, r_{12}, \>
        r_9 \, r_{12}, \>
        r_1 \, r_{13}, \>
        r_5 r_{13}, \>
        r_6 r_{13}, \>
        r_1 r_{14}, \>
        r_5 r_{14}, \>
        r_9 r_{14}
    }
    \>.
    \end{aligned}
    \label{eq:quad_charges}
\end{align}
Together, they give a total of 46 non-trivial charges.
For example, an object with quadratic charge, say $r_i r_j$, is odd w.r.t.\ the sign of $r_i$, of $r_j$, and of the product $r_i r_j$, but is even w.r.t.\ flipping the signs of both $r_i$ and $r_j$ simultaneously.
This organisation allows us to break down the problem into smaller pieces:
subsets of connection-matrix elements with different charges cannot be $\mathbb{Q}$-linearly related to each other.

Next, we focus on the matrix elements that have a given charge, say $\mathcal{C}$.
We determine the possible $\dlog W_{\alpha}(\vec{x})$'s with charge $\mathcal{C}$ by using the \soft{Mathematica} package \soft{BaikovLetter}~\cite{Jiang:2024eaj}.
If $\mathcal{C}=1$, the letters $W_{\alpha}(\vec{x})$ are rational, and we complement \soft{BaikovLetter}'s output with the irreducible denominator factors read from the connection matrix.
If instead $\mathcal{C}$ is a non-trivial product of square roots, the corresponding odd letters $W_{\alpha}(\vec{x})$ are algebraic and take the form
\begin{align}
  W(\vec{x}) = \frac{A + B \> \cC}{A - B \> \cC}
    \>,
\end{align}
where $A$ and $B$ are polynomials in $\vec{x}$ of degree up to 4.
We then derive a basis of the vector space over $\mathbb{Q}$ spanned by all connection matrix elements and $\dlog$'s with charge $\mathcal{C}$.
In doing this, we prefer the $\dlog$'s over the connection matrix elements, and choose the residual non-$\dlog$ one-forms $\omega_{\beta}(\vec{x})$ in the basis so that they are as compact as possible.
By iterating this over all possible charges and rewriting the connection matrices in terms of the resulting bases of one-forms, we obtain \cref{eq:DE_dlog_one_form}.
We summarise the key features of the resulting DEs in \cref{tab:dlog_one_form_stats}.

For families $F_1$, $F_2$, and $F_3$ combined,
the alphabet of letters $W_\alpha$ appearing in
\cref{eq:DE_dlog_one_form} consists of 164 entries: 65 rational and 99 algebraic.
The rational letters are polynomials in $\vec{x}$, with total degree up to $5$ for family $F_1$.
For families $F_2$ and $F_3$, we also find additional
contributions from polynomials of degree 9 and 14.
The former is spurious and could be removed by choosing different MIs, while the latter is related to the discriminant of the five-point
elliptic curve; see \cref{sec:five-point_elliptic,app:elliptic_details} for more details.

Three non-$\dlog$ one-forms, appearing only in the DEs for $F_1$, are closed and exact.\footnote{The closed one-forms of $F_1$ are denoted by \texttt{w[F1,r[1],\brc{10,11},0]}, \texttt{w[F1,r[6],\brc{34, 35},1]}, and \texttt{w[F1,r[7],\brc{31,32},1]} in our ancillary files~\cite{zenodo}.}
Two of them have an algebraic primitive,
\begin{align}
\dd A^{(F_1)}_{34, 35}(\vec{x}) \Bigl|_{\eps^1} & = \dd \left[ \frac{4 m_t^2}{s_{45} (m_t^2-s_{15})} \ r_6(\vec{x}) \right] \,, \\
\dd A^{(F_1)}_{31, 32}(\vec{x}) \Bigl|_{\eps^1} & = \dd \left[ \frac{4 m_t^2}{s_{12} (m_t^2-s_{15})} \ r_7(\vec{x}) \right] \,.
\end{align}
The third one, giving the connection matrix elements $\dd A^{(F_1)}_{10,11}$ and $\dd A^{(F_1)}_{11,10}$ at order $\eps^0$, can be integrated in terms of logarithms, but their arguments contain nested square roots;
therefore, we do not to make use of a $\dlog$ representation for it.
The closedness of these one-forms has no impact on the work presented here, but strengthens the analogy with the elliptic sector studied in refs.~\cite{Badger:2024fgb,Becchetti:2025oyb}, where a similar phenomenon was observed~\cite{Badger:2024dxo}.

\begin{table}[t]
    \centering
    \begin{tabular}{ccccccc}
        \toprule
        family & basis size & elliptic curves & \begin{tabular}{@{}c@{}} nested \\ square roots \end{tabular} & entries & letters & \begin{tabular}{@{}c@{}} non-$\dlog$ \\ one-forms \end{tabular} \\
        \midrule
        $F_1$ & 141 & 2 (\cref{fig:ttj_elliptic,fig:4pt_elliptic}) & 1 (\cref{fig:nested}) & 2339 & 101 & 119 \\
        $F_2$ & 122& 1 (\cref{fig:5pt_elliptic}) & 0 & 2027 & 122 & 84 \\
        $F_3$ & 131 & 1 (\cref{fig:5pt_elliptic}) & 0 & 2333 & 137 & 96 \\
        \bottomrule
    \end{tabular}
    \caption{
       Numbers of MIs, of sectors with elliptic curves, of sectors with nested square roots, of non-zero entries of the connection matrix, of letters and of non-$\dlog$ one-forms in \cref{eq:DE_dlog_one_form} for the three integral families.
    }
    \label{tab:dlog_one_form_stats}
\end{table}

The analytic expression of the non-$\dlog$ one-forms is considerably bulkier than in other similar cases in the literature~\cite{FebresCordero:2023pww,Badger:2024fgb}, in particular for $F_2$ and $F_3$.
For example, the most complicated irreducible polynomial, appearing in the DEs for families $F_2$ and $F_3$, has total degree $35$ and counts $556685$ monomials in the $7$ variables $\vec{x}$.
In order to optimise the expressions and make their evaluation more efficient, we collect the common irreducible polynomial factors across all one-forms of each family, and optimise the polynomials with an ad-hoc mix of \soft{FORM}~\cite{Ruijl:2017dtg} (format \texttt{O4}) and \soft{Mathematica}.

\subsection{Numerical evaluation of the solution}
\label{sec:Evaluations}

In this section we apply the multivariate generalisation of the method
of generalised power series expansions~\cite{Pozzorini:2005ff,Aglietti:2007as,Lee:2017qql,Lee:2018ojn,Bonciani:2018uvv,Fael:2021kyg,Fael:2022rgm}
proposed in ref.~\cite{Francesco:2019yqt} to evaluate numerically the solution to the DEs derived in this work.
A number of public implementations of this method are by now available~\cite{Hidding:2020ytt,Liu:2022chg,Armadillo:2022ugh,Prisco:2025wqs}; we make use of the \soft{Mathematica} package \soft{DiffExp}~\cite{Hidding:2020ytt}.

In this approach, the DEs are integrated along a univariate path connecting an initial phase-space point, where the values of the MIs are known, to the desired target point.
For simplicity, this path is chosen to be a straight line.
In order to avoid subtleties related to the analytic continuation, we restrict to the physical channel relevant for $t\bar{t}W$ production ($35 \to 124$).
The latter is defined by the following inequalities,
\begin{align}
\begin{aligned}
p_3 \cdot p_1 \,, \ \ p_3 \cdot p_2 \,, \ \ p_3 \cdot p_4 \,, \ \ p_5 \cdot p_1 \,, \ \ p_5 \cdot p_2 \,, \ \ p_5 \cdot p_4 < 0 \,, \\
p_3 \cdot p_5\,,  \ \ p_1\cdot p_2\,, \ \ p_1\cdot p_4\,, \ \ p_2\cdot p_4 > 0\,, \\
p_1^2 = p_2^2 > 0 \,, \\
\ p_4^2 > 0 \,,
\end{aligned}
\end{align}
complemented by the following Gram-determinant constraints,
\begin{align}
 G(p_i, p_j) < 0 \,, \qquad G(p_i,p_j,p_k) > 0 \,, \qquad G(p_1,p_2,p_3,p_4) < 0 \,,
\end{align}
for all distinct $i, j, k \in \{1,\ldots,5\}$.
We take the square roots of negative arguments to have positive imaginary part.

We obtained numerical values of the MIs up to order $\eps^4$ at $10$ phase-space points with at least $35$-digit precision by means of the \soft{Mathematica} package \soft{AMFlow}~\cite{Liu:2022chg}, which implements the method of the auxiliary mass flow~\cite{Liu:2017jxz, Liu:2021wks}.
The points are inside the physical region, and are chosen so that the values of the invariants $\vec{x}$ are ratios of small integers.
This speeds up the evaluation with \soft{AMFlow}.
For example, one of the chosen points is
\begin{align}
\label{eq:X1}
\vec{x}_1 = \left\{ 5, - \frac{3}{10}, - \frac{9}{2}, -\frac{8}{3}, - \frac{7}{10}, \frac{1}{3}, 1 \right\} \,.
\end{align}
By comparing the values of the MIs at the $10$ points, we verified that
\begin{itemize}
 \item they are constant and rational at order $\eps^0$;
 \item the imaginary part is given by $\ii \pi$ times rational constants at order $\eps^1$;
 \item the real part is given by $\mathbb{Q}$-linear combinations of the expected logarithms at order~$\eps^1$.
\end{itemize}
The expected logarithms in the last item correspond to the allowed branch cuts, and have been determined in ref.~\cite{Becchetti:2025osw}.
These properties are expected for MIs that satisfy canonical DEs;
the fact that they are satisfied by our bases is a sign that we have successfully `pushed' the non-canonical features to higher orders in $\eps$.

Given the high degree of algebraic complexity of the DEs for families $F_2$ and $F_3$, the pull back of the DEs to the univariate path takes a considerable amount of time.
If done by \soft{DiffExp}, this step often exceeds the time required to evolve the solution from the initial to the target point.
For this reason, we carry out this step with a separate \soft{Mathematica} script, which leverages the optimised form of the DEs discussed in \cref{sec:DEs_Form}, achieving a speed up by an order of magnitude.
Then, we resort to \soft{DiffExp} only for the solution of the DEs on the path.

We validated our results by integrating the DEs up to order $\eps^4$ between various pairs of points from the $10$ at which we ran \soft{AMFlow}.
Each of the $10$ points was used as a target point at least once.
The pairs are chosen so that the line connecting them stays within the physical channel.\footnote{If the line between the initial and the target point leaves the physical region, one can change the starting point, use a piece-wise straight line, or derive the prescriptions for the analytic continuation. The latter option however requires more work, and the evolution of the solution across physical singularities is typically more time consuming.}
In all cases we found agreement between the \soft{DiffExp} and \soft{AMFlow} results within the \soft{DiffExp} target precision (25 digits).

We stress that these tests are only meant to validate our results.
There are several promising avenues to make our results more suitable for direct usage in phenomenology, acting on both the expression of the DEs and the solution strategy.
Cases in point are the use of multivariate partial fractioning to further reduce the maximal degree of the appearing polynomials, the use of the recently published \soft{C} implementation of the generalised series expansion method~\cite{Prisco:2025wqs}, and the construction of a basis of special functions to span the solution up to the required order in $\eps$ as proposed in ref.~\cite{Badger:2024dxo}.
Moreover, the evaluation time depends strongly on the segmentation of the integration path, which in turn depends on the end-points and on the nearby singularities.
An evaluation strategy aimed at a large number of phase-space points should therefore also minimise the total number of segments, which can be done by re-using iteratively the targets of previous evaluations as initial points.
We leave the study of all these optimisations to future work.

We provide the numerical values of all MIs at the $10$ phase-space points, the definition of the physical region, and a \soft{Mathematica} script to parametrise the DEs on a line and solve them with \soft{DiffExp} in our ancillary files~\cite{zenodo}.

\section{Conclusion}
\label{Conclusions}

In this paper, we take significant steps towards the exact computation of the two-loop scattering amplitudes relevant for the NNLO QCD corrections to $t\bar{t}W$ production at hadron colliders.
Specifically, we evaluate a complete set of master integrals of the three two-loop five-point Feynman integral families shown in \cref{fig:Families}.
All Feynman integrals appearing in the two-loop amplitude in the leading colour approximation can be reduced to the latter or to products of the one-loop integrals computed in ref.~\cite{Becchetti:2025osw}.
We identify three elliptic curves that underlie the geometry of the integrals under consideration, as well as nested square roots in their algebraic structure.
These analytic properties prevent us from applying the established methodology that underpins all NNLO QCD computations for $2\to 3$ processes obtained so far.
We construct integral bases obeying DEs that depend at most quadratically on the dimensional regulator for sectors where these analytic structures appear, and are canonical otherwise.
We represent the DEs in an optimised way that minimises repeated patterns and separates clearly the canonical from the non-canonical features.
This allows us to evaluate the master integrals with the public implementations of the method of generalised power series~\cite{Moriello:2019yhu}.
We provide benchmark evaluations by means of the \soft{Mathematica} package \soft{DiffExp}~\cite{Hidding:2020ytt}.
Our results open several interesting avenues.

A natural next step is the construction of canonical DEs.
Building on the method of ref.~\cite{Gorges:2023zgv}, the authors of ref.~\cite{Becchetti:2025oyb} have recently obtained canonical DEs for a two-loop five-point integral family for the first time.
This development suggests that the methodology may be mature enough to tackle high-multiplicity processes.
The increased algebraic complexity and the presence of multiple elliptic curves make the integrals considered in this work ideal to stress-test and further refine the method of ref.~\cite{Gorges:2023zgv}.
Indeed, as shown in refs.~\cite{Badger:2024fgb,Becchetti:2025oyb} for $t\bar{t}j$ production, the form of the DEs obtained here is a perfect starting point to apply this method.

It would also be of great interest to find ways to handle the solution that do not rely on the canonical form but can still be useful for phenomenology.
An alternative approach going in this direction was proposed recently in ref.~\cite{Badger:2024dxo}, and 
our results appear to be well suited to be treated the same way.
This would lead to a simpler representation of the two-loop amplitudes, in which the $1/\eps$ poles can be cancelled analytically, although more study will still be required to make their numerical evaluation suitable for phenomenological applications.

Finally, the integral bases we constructed and the information we gathered about their analytic structure open the door to an efficient computation of the two-loop amplitudes in the leading colour approximation for this important LHC process.

\acknowledgments

We would like to thank Simon Badger, Maximilian Delto, Sara Ditsch, Phillip Kreer and Lorenzo Tancredi for useful discussions and comments on the draft. This work was supported by the European Research Council (ERC) under the European Union's Horizon Europe research and innovation program grant agreement 101040760, \textit{High-precision multi-leg Higgs and top physics with finite fields} (ERC Starting Grant \emph{FFHiggsTop}), and by the Swiss National Science Foundation (SNSF) under the Ambizione grant No.~215960.

\appendix

\section{Description of the ancillary files}
\label{Ancillary}

We list and describe here the content of our ancillary files, available at ref.~\cite{zenodo}.
All files contain expressions in \soft{Wolfram Mathematica} language.
The notation is the following.
\begin{itemize}
\item Families: $F_1 = $~\code{F1}, $F_2 = $~\code{F2}, $F_3 = $~\code{F3}.
\item Variables: $s_{ij}=$\code{sij}, $m_t^2=$~\code{mt2}, $m_t=$~\code{mt}, $m_w^2=$~\code{mw2}, $m_w=$~\code{mw}, $\eps=$~\code{eps}.
\item Square and nested square roots: $r_i=$~\code{r[i]}, $\mathrm{NS}_{+} = $~\code{NS[``+'']}, $\mathrm{NS}_{-} = $~\code{NS[``-'']}.
\item Feynman integrals: $I_{ijk\cdots}^{(F)} =$~\code{j[F,\{i,j,k,...\}]}.
\item Momenta: $p_i =$~\code{pi} (for $i=1,\ldots,5$), $k_1 =$~\code{k1}, $k_2 =$~\code{k2}.
\item Other symbols: $\dlog W =$~\code{dlog[W]}, $\omega = $~\code{w}, $\mu_{ij} = $~\code{muij}.
\end{itemize}

\smallskip
\noindent
We now list the files and describe their content.
\begin{itemize}
 \item \code{square\_roots.m} -- List of replacement rules defining the \code{r[i]}'s as square roots (\code{sqrt[...]}) of polynomials in $\vec{x}$ (cfr.~\cref{eq:roots1,eq:roots2}).

 \item \code{nested\_square\_roots.m} -- List of replacement rules defining the nested square roots \code{NS[``+'']} and \code{NS[``-'']} (cfr.~\cref{eq:nested_sqrt}).

 \item \code{alphabet.m} -- List of replacement rules defining the letters, i.e., the arguments of the $\dlog$ one-forms (cfr.~\cref{sec:DEs_Form}).
   We denote each letter as \code{W[R,i]}, where \code{i} is an integer and \code{R} is the charge of the letter.
   E.g., \code{R} is \code{1} for a rational letter, and \code{r[i]} for an algebraic letter that is odd w.r.t.\ $r_i$.

 \item \code{physical\_region.m} -- List of inequalities that have to be satisfied for a phase-space point $\vec{x}$ to be in the physical region defined in \cref{sec:Evaluations}.

 \item \code{solve.wl} -- \soft{Mathematica} script that evaluates the MIs up to order $\eps^4$ by solving the corresponding DEs (cfr.~\cref{sec:Evaluations}).
   It relies on \soft{DiffExp}~\cite{Hidding:2020ytt}.
   The user must specify the path to the file \code{DiffExp.m} (\code{\$PathToDiffExp}), the family (\code{family}~$=$~\code{F1}, \code{F2} or \code{F3}), and the initial/final point (\code{XinLabel}/\code{XfinLabel}).
   Initial and final points can be chosen amongst the $10$ (\code{X1}, \code{X2}, ..., \code{X10}) at which we obtained values with \soft{AMFlow}~\cite{Liu:2022chg}.
   The target accuracy is set to $20$ digits.
   The script checks that the integration path lies within the physical region, and that the results are in agreement with the benchmark values obtained with \soft{AMFlow}.
   All provided points are connected to \code{X1} by a straight line in the physical region.
\end{itemize}

\smallskip
\noindent
For each family \code{F}~$=$~\code{F1}, \code{F2}, \code{F3}, there is a folder \code{F/} containing the following files:
\begin{itemize}
	\item
      \code{F\_propagators.m} -- List of generalised inverse propagators $D_{F,i}$ with $i=1,\ldots,11$ (cfr.~\cref{tab:propagators}).

    \item
      \code{F\_basis.m} -- List of MIs expressed in terms of scalar Feynman integrals and factors of \code{muij} (cfr.~\cref{sec:Bases_and_DEs}),
      without the square-root normalisation.
      The factors of \code{muij} depend on the loop momenta (cfr.~\cref{eq:muij}), and have to be taken under the integral sign of the scalar integral they multiply.

    \item
        \code{F\_basis\_expanded.m} -- List of MIs expressed in terms of scalar Feynman integrals, without the square-root normalisation.

    \item
        \code{F\_sqrt\_normalisation.m} -- List of products of square roots that give the normalisation of the MIs.
           The complete expression of the $i$th MI, $\mathrm{I}_i^F$, is obtained by multiplying the $i$th entry of \code{F\_sqrt\_normalisation.m} by the $i$th entry of \code{F\_basis.m} (or \code{F\_basis\_expanded.m}).

  \item
        \code{F\_connection\_matrix.m} -- Connection matrix $\dd A^{(F)}$ given in terms of $\dlog$ (\code{dlog[W]}, with the letters $W$ defined in \code{alphabet.m}) and non-$\dlog$ one-forms (\code{w}, defined in the next file), and \code{eps} (cfr.~\cref{eq:DE_dlog_one_form}).

   \item
        \code{F\_one-forms.m} -- Definition of the non-$\dlog$ one-forms $\omega_i$ of family \code{F} as a list of two elements.
        The first element is a list of replacement rules giving the expression of each one-form as
        \begin{align*}
          \code{w[F,R,\{i,j\},k] -> \{c1,c2,\ldots,c7\}} \,,
        \end{align*}
        where \code{c1}, ..., \code{c7} are the coefficients of the differentials of the variables $\vec{x}$, written in terms of square roots and irreducible polynomials \code{y[k]}.
        The arguments of \code{w} keep track of its charge \code{R} (either \code{1} or a product of square roots), and of the entry of the connection matrix chosen to define it.
        For example, \code{w[F,R,\{i,j\},k]} equals the coefficient of $\eps^k$ in $\dd A^{(F)}_{ij}$.
        The second element is a list of replacement rules defining the \code{y[k]}'s as polynomials in $\vec{x}$.

  \item
        \code{values/F\_Xi.m} (\code{i}~$ = 1, \ldots, 10$) -- Values of the MIs at the phase-space point \code{Xi} obtained with \soft{AMFLow} with (at least) $35$-digit precision.
        The first element of the file defines the phase-space point as a list of replacement rules for the invariants $\vec{x}$.
        The second element gives the numerical values of the MIs up to order $\eps^4$.
\end{itemize}

\smallskip
\noindent
Finally, the folder \code{elliptic\_curves/} contains the following files related to the elliptic curves:
\begin{itemize}
	\item 
	    \code{4pt\_4var\_curve\_quartic.m}, \code{4pt\_4var\_j-invariant.m} -- Expressions of the quartic polynomial in \code{z} and of the $j$-invariant related to the elliptic curve
	    appearing in the four-point sector shown in \cref{fig:ttj_elliptic} of $F_1$ (cfr.~\cref{sec:Elliptic_curves}).
	\item	   
	    \code{4pt\_5var\_curve\_quartic.m}, \code{4pt\_5var\_j-invariant.m} -- Expressions of the quartic polynomial $\mathcal{P}_{4\text{-pt.}}$ in \code{z} (cfr.~\cref{eq:P4pt}) and of the $j$-invariant related to the elliptic curve
	    appearing in the four-point sector shown in \cref{fig:4pt_elliptic} of $F_1$ (cfr.~\cref{sec:four-point_elliptic}).
	\item
	   \code{5pt\_7var\_curve\_cubic.m}, \code{5pt\_7var\_curve\_quartic.m}, \code{5pt\_7var\_j-invariant.m} --
       Respectively, the Weierstra\ss{} cubic polynomial $\mathcal{P}^{\rm W.}_{5\text{-pt.}}$ in \code{z} (cfr.~\cref{eq:P5ptW}),
	   the quartic polynomial $\mathcal{P}_{5\text{-pt.}}$ in \code{z} (cfr.~\cref{eq:elliptic_monster}),
	   and the $j$-invariant $j_{5\text{-pt.}}$ (cfr.~\cref{eq:jinv_5pt})
	   related to the elliptic curve appearing in the five-point sector of $F_2$ and $F_3$ shown in \cref{fig:5pt_elliptic} (cfr.~\cref{sec:five-point_elliptic,app:elliptic_details}).
\end{itemize}

\section{Master integrals for the five-point sectors}
\label{New_Sectors}

In this appendix we collect the MIs for the five-point sectors, excluding those already presented in \cref{sec:Nested_square_roots} (associated with the nested square roots) and \cref{sec:five-point_elliptic} (associated with the five-point elliptic curve).
Note that there are also four-point sectors that have not been computed before; we omit them from this appendix and refer the interested readers to the ancillary files~\cite{zenodo}.
We also stress that the MIs that do not couple to the elliptic sectors (discussed in \cref{sec:Elliptic_curves}) are pure, i.e., they satisfy canonical DEs with solely $\dlog$ one-forms.

We label the sectors by the indices of the appearing propagators: $(\nu_1,\ldots,\nu_{11})$, where $\nu_i=1$ if the $i$th (generalised) inverse propagator is present in the denominator, and $\nu_i=0$ if it only appears in the numerator.
We recall that $\nu_{9}, \nu_{10}, \nu_{11} = 0$ for all sectors, as they correspond to irreducible scalar products, and that the propagators are defined in \cref{tab:propagators}.
For each sector, we give the corresponding Feynman graph and a list of numerators defining the MIs.
For a given numerator $N^{(F)}_i$, the corresponding MI is $I_{\vec{\nu}}^{(F)}[N]$, where $\vec{\nu}$ are the propagator indices defining the sector, and the subscript $i$ gives the position of the MI in the basis.
We present only once, for an arbitrary family, the sectors that appear in multiple families.

\begin{samepage}

\subsection*{Eight-propagator sectors}

\begin{center}
  \textbf{$F_{1}$, $(1,1,1,1,1,1,1,1,0,0,0)$, 3 MIs}
\end{center}
\vspace{-0.5cm}
\hspace{\parindent}
\begin{minipage}{0.37\textwidth}
  \begin{figure}[H]
    \centering
    \includegraphics[width=\textwidth]{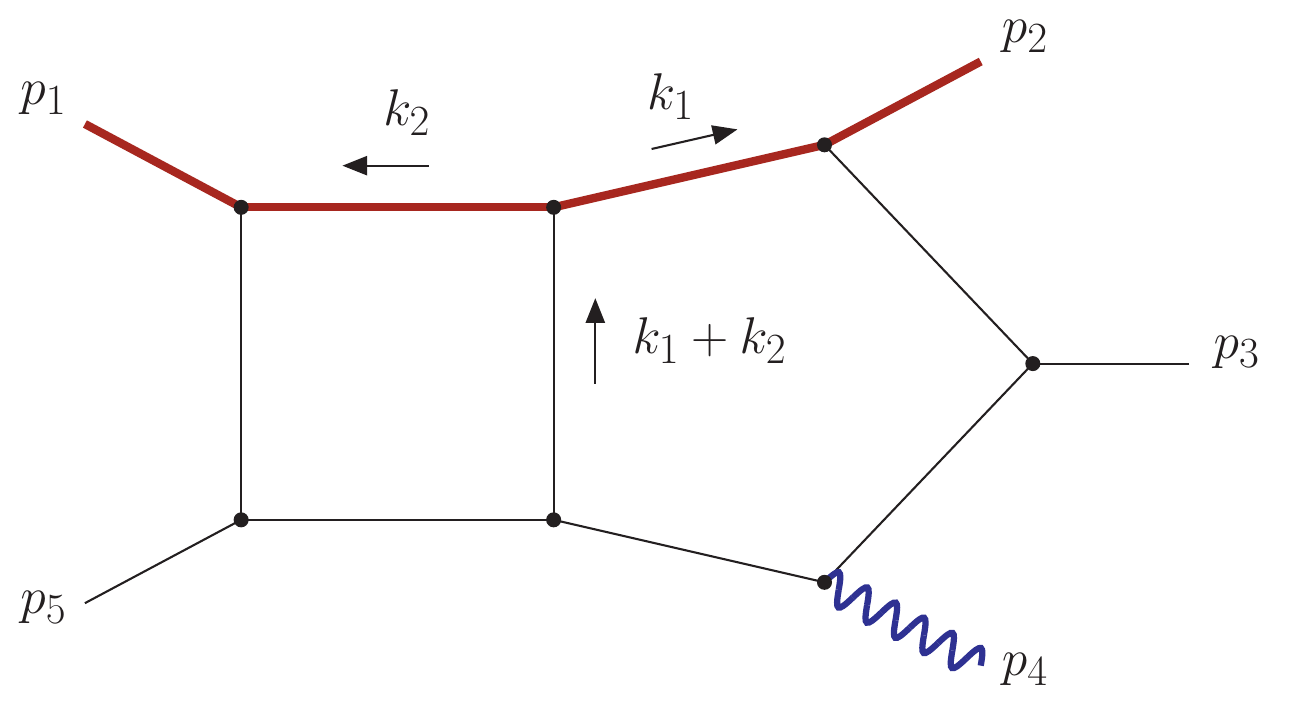}
  \end{figure}
\end{minipage}%
\hspace{0.02\textwidth}
\begin{minipage}{0.61\textwidth}
  $\begin{aligned}
    N_1^{(F_1)}&= \eps^4 r_1 (s_{15} - m_t^2) \mu_{11} \,, \\
    N_2^{(F_1)}&= \eps^4 r_1 (s_{15} - m_t^2) \mu_{12} \,, \\
    N_3^{(F_1)}&= \eps^4 s_{34}(m_t^2 - s_{15}) (m_t^2 - s_{23}) (m_t^2 + D_{F_1,9}) \,.
  \end{aligned}$
\end{minipage}

\end{samepage}

\medskip

\begin{samepage}

\begin{center}
  \textbf{$F_{2}$, $(1,1,1,1,1,1,1,1,0,0,0)$, 3 MIs}
\end{center}
\vspace{-0.5cm}
\hspace{\parindent}
\begin{minipage}{0.37\textwidth}
  \begin{figure}[H]
    \centering
   \includegraphics[width=\textwidth]{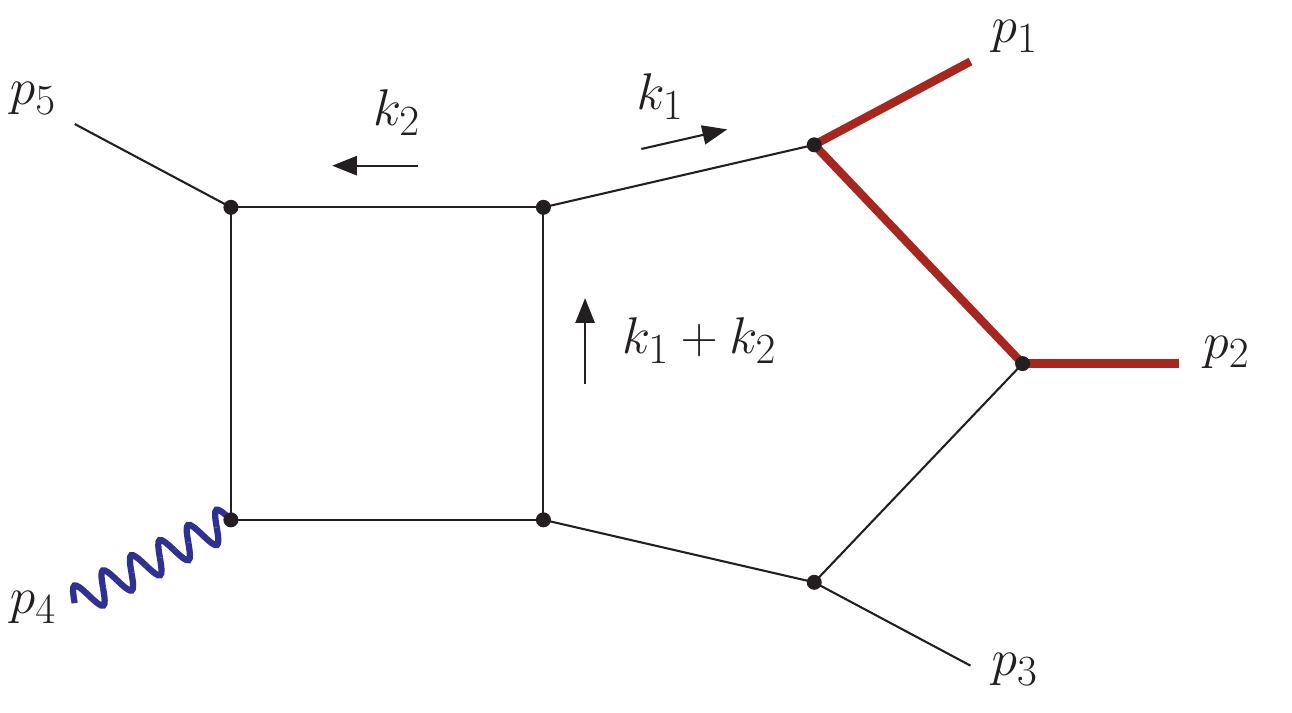}
  \end{figure}
\end{minipage}%
\hspace{0.02\textwidth}
\begin{minipage}{0.61\textwidth}
  $\begin{aligned}
    N_1^{(F_2)}&= \eps^4 r_1 s_{45} \mu_{11} \,, \\
    N_2^{(F_2)}&= \eps^4 r_1 s_{45} \mu_{12} \,, \\
    N_3^{(F_2)}&= \eps^4  s_{12} \left(m_t^2 - s_{23}\right) \left(m_w^2 D_{F_2,1}-s_{45} D_{F_2,9}\right) \, .
  \end{aligned}$
\end{minipage}

\end{samepage}

\medskip

\begin{samepage}

\begin{center}
  \textbf{$F_{3}$, $(1,1,1,1,1,1,1,1,0,0,0)$, 4 MIs}
\end{center}
\vspace{-0.5cm}
\hspace{\parindent}
\begin{minipage}{0.37\textwidth}
  \begin{figure}[H]
    \centering
    \includegraphics[width=\textwidth]{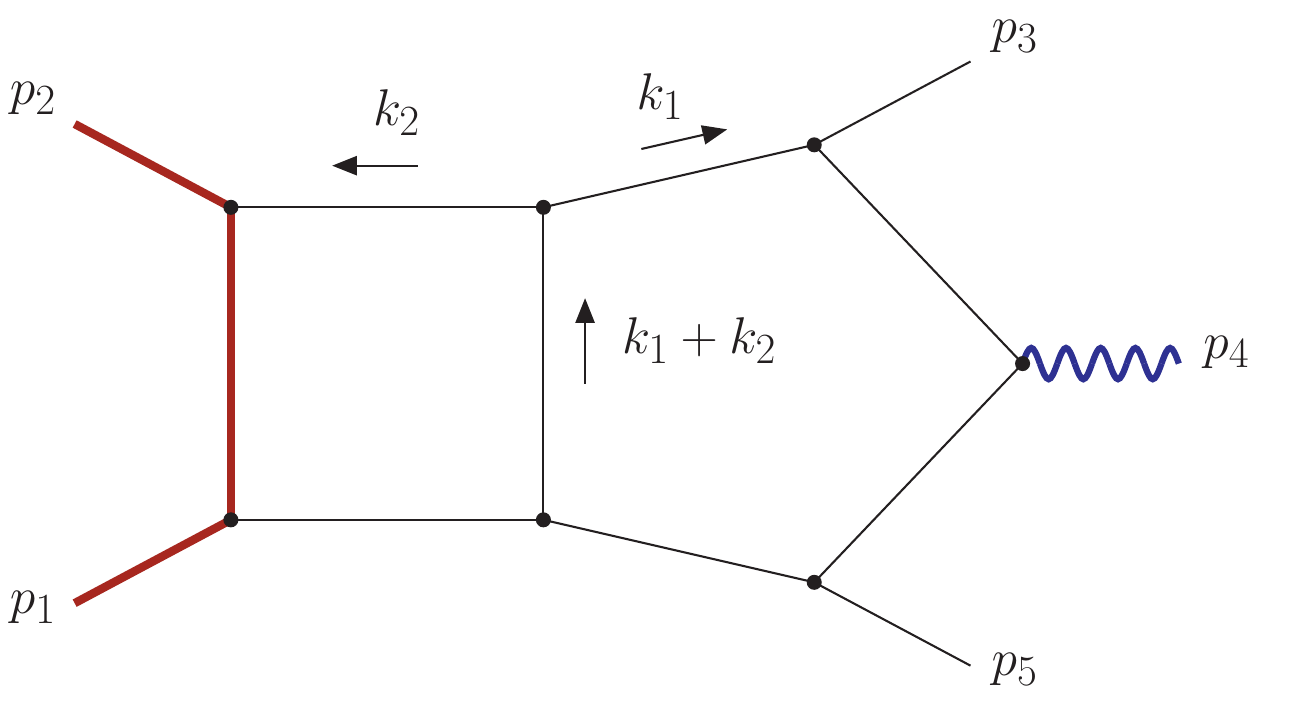}
  \end{figure}
\end{minipage}%
\hspace{0.02\textwidth}
\begin{minipage}{0.61\textwidth}
  $\begin{aligned}
    N_1^{(F_3)}&= \eps^4 r_1 r_9 \left(\frac{2-\eps}{2+\eps} \frac{\mu_{12}^2-\mu_{11} \mu_{12}}{s_{12}}-\frac{\mu_{11}+\mu_{12}}{2}\right) \,, \\
    N_2^{(F_3)}&= \eps^4 r_1 s_{12} \mu_{11} \,, \\
    N_3^{(F_3)}&= \eps^4 r_1 s_{12} \mu_{12} \,, \\
    N_4^{(F_3)}&= \eps^4 s_{12} \left(m_w^2 s_{12} -s_{34} s_{45}\right) D_{F_3,9} \,.
  \end{aligned}$
\end{minipage}

\end{samepage}

\bigskip

\begin{samepage}

\subsection*{Seven-propagator sectors}

\begin{center}
  \textbf{$F_{2}$, $(1,1,1,1,0,1,1,1,0,0,0)$, 1 MI}
\end{center}
\vspace{-0.5cm}
\hspace{\parindent}
\begin{minipage}{0.37\textwidth}
  \begin{figure}[H]
    \centering
    \includegraphics[width=\textwidth]{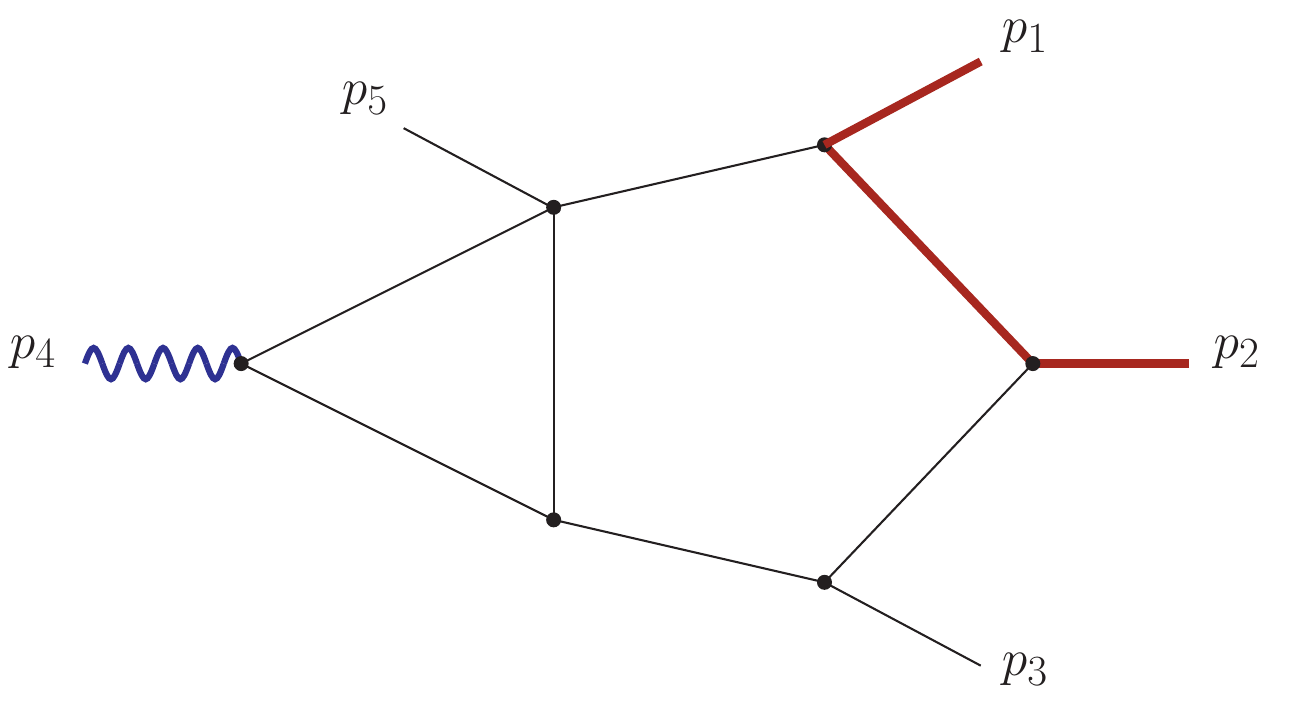}
  \end{figure}
\end{minipage}%
\hspace{0.02\textwidth}
\begin{minipage}{0.61\textwidth}
  $\begin{aligned}
    N_{16}^{(F_{2})}&= \eps^4 r_1 \mu_{11} \,.
  \end{aligned}$
\end{minipage}

\end{samepage}

\medskip

\begin{samepage}

\begin{center}
  \textbf{$F_{2}$, $(0,1,1,1,1,1,1,1,0,0,0)$, 3 MIs}
\end{center}
\vspace{-0.5cm}
\hspace{\parindent}
\begin{minipage}{0.37\textwidth}
  \begin{figure}[H]
    \centering
    \includegraphics[width=\textwidth]{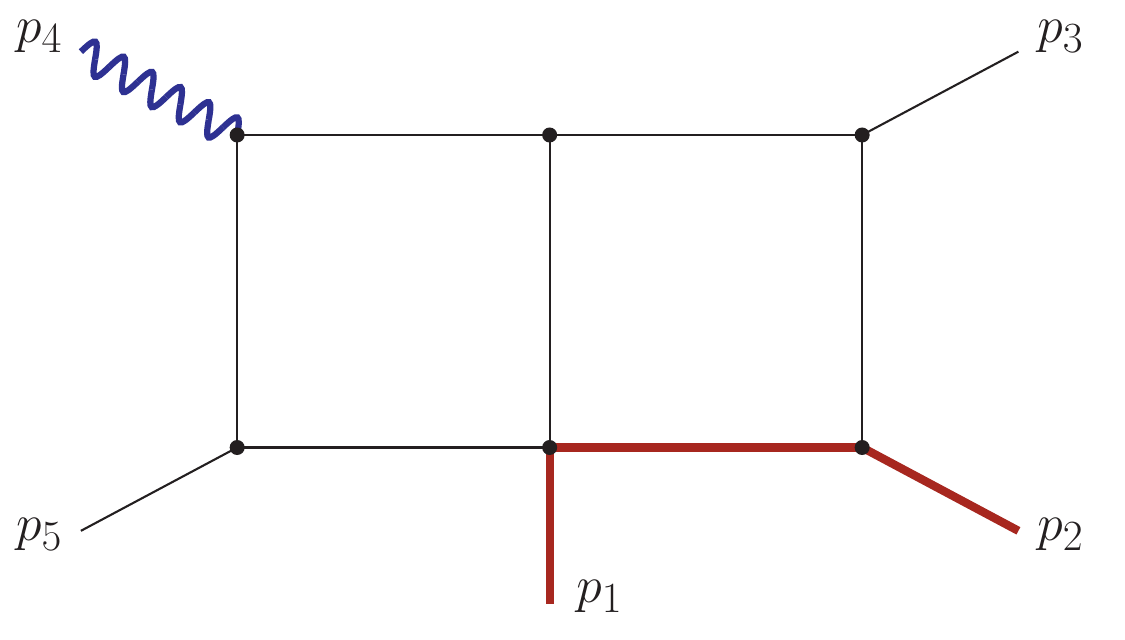}
  \end{figure}
\end{minipage}%
\hspace{0.02\textwidth}
\begin{minipage}{0.61\textwidth}
  $\begin{aligned}
    N_{11}^{(F_{2})}&= \eps^4 r_1 \mu_{12} \,, \\
    N_{12}^{(F_{2})}&= \eps^4 (m_t^2 - s_{23})\left(m_w^2 - s_{45}\right) D_{F_2,11} \,, \\
    N_{13}^{(F_{2})}&= \eps^4 (m_t^2 - s_{23}) \left(m_w^2 s_{12} - s_{34} s_{45}\right) \,.
  \end{aligned}$
\end{minipage}

\end{samepage}

\medskip

\begin{samepage}

\begin{center}
  \textbf{$F_{2}$, $(1,1,1,0,1,1,1,1,0,0,0)$, 4 MIs}
\end{center}
\hspace{\parindent}
\begin{minipage}{0.37\textwidth}
  \begin{figure}[H]
    \centering
    \includegraphics[width=\textwidth]{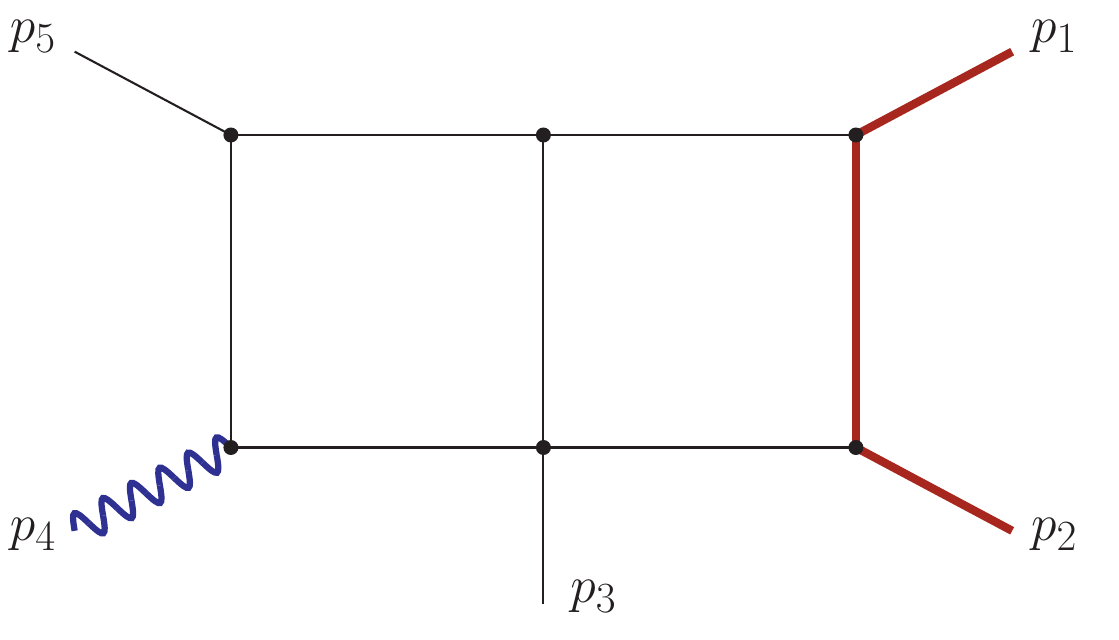}
  \end{figure}
\end{minipage}%
\hspace{0.02\textwidth}
\begin{minipage}{0.61\textwidth}
  $\begin{aligned}
    N_{4}^{(F_{2})}&= \eps^4 r_1 \mu_{12} \, ,\\
    N_{5}^{(F_{2})}&= \eps^4 s_{12} \bigl[ \left(m_w^2 - s_{45}\right) D_{F_2,10} \\
      & \phantom{=} \ +  s_{45} \left(m_t^2 - s_{15}\right)\bigr] \,, \\
    N_{6}^{(F_{2})}&= \eps^4 r_9 \left(s_{45} D_{F_2,9} - m_w^2 D_{F_2,1}\right) \,,\\
    N_{7}^{(F_{2})}&= \eps^4 s_{12} s_{45} \left(s_{15}-m_t^2\right) \,.
  \end{aligned}$
\end{minipage}

\end{samepage}

\medskip

\begin{samepage}

\begin{center}
  \textbf{$F_{1}$, $(1,1,1,0,1,1,1,1,0,0,0)$, 6 MIs}
\end{center}
\hspace{\parindent}
\begin{minipage}{0.37\textwidth}
  \begin{figure}[H]
    \centering
    \includegraphics[width=\textwidth]{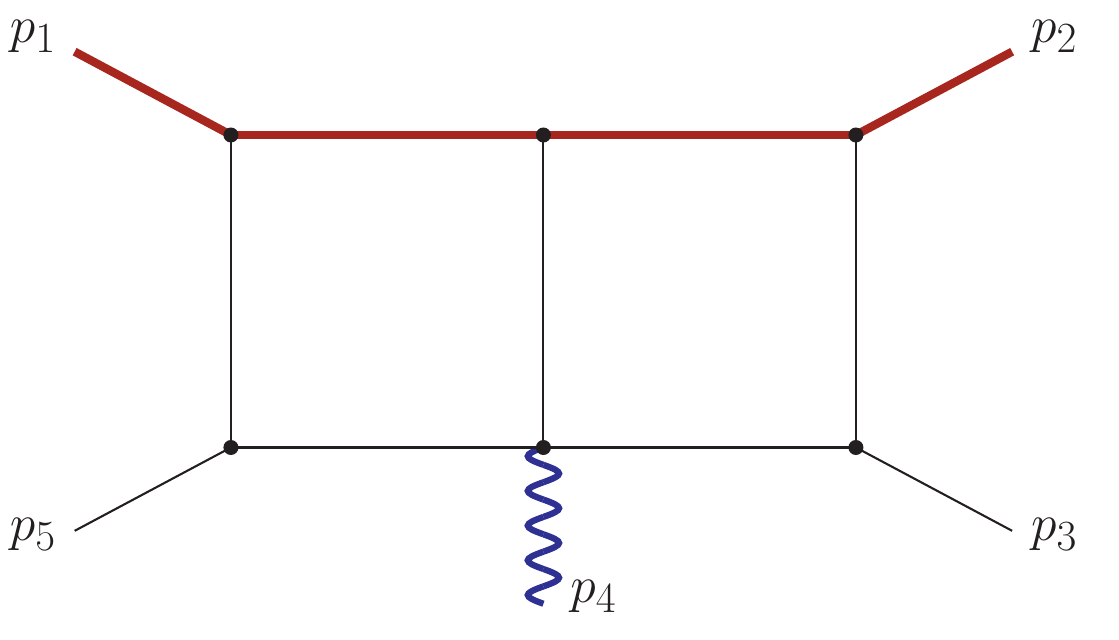}
  \end{figure}
\end{minipage}%
\hspace{0.02\textwidth}
\begin{minipage}{0.61\textwidth}
  $\begin{aligned}
    N_5^{(F_{1})}&= \eps^3 \biggl[\eps \left(f_1 D_{F_1,10} + f_2 \right)+f_3  \\
      & \phantom{=} \ + \frac{f_4 D_{F_1,11} + f_5 D_{F_1,10}+ f_6}{D_{F_1,7}} \\
      & \phantom{=} \ + \mu_{12} \left( \frac{f_7}{D_{F_1,3}}+\frac{f_8}{D_{F_1,7}}\right) \biggr] \\
      & \phantom{=} \ +\text{(sub-sectors)} \,,\\
    N_6^{(F_{1})}&= \eps^3 r_1 (s_{15} - m_t^2) \frac{\mu_{12} }{ D_{F_1,7}} \,, \\
    N_7^{(F_{1})}&= \eps^3 r_1 (s_{23} - m_t^2) \frac{\mu_{12} }{ D_{F_1,3}} \,, \\
    N_8^{(F_{1})}&= \eps^4 r_1 \mu_{12} \, , \\
    N_9^{(F_{1})}&= \eps^4 \left(m_t^2 - s_{15}\right) \left(m_t^2 - s_{23}\right) \left(m_t^2 + D_{F_1,10}\right) \,,\\
    N_{10}^{(F_{1})}&= \eps^4 s_{12} \left(m_t^2 - s_{15}\right) \left(m_t^2 - s_{23}\right) \,,
  \end{aligned}$
\end{minipage}

\vspace{0.2cm}
\noindent
where $f_1$, \ldots, $f_8$ are rational functions of the kinematic invariants $\vec{x}$, and we omit the contributions from the sub-sectors.

\end{samepage}

\bigskip

\subsection*{Six-propagator sectors}

\begin{samepage}

\begin{center}
  \textbf{$F_{2}$, $(0,1,1,1,1,1,0,1,0,0,0)$, 2 MIs}
\end{center}
\vspace{-0.5cm}
\hspace{\parindent}
\begin{minipage}{0.37\textwidth}
  \begin{figure}[H]
    \centering
    \includegraphics[width=\textwidth]{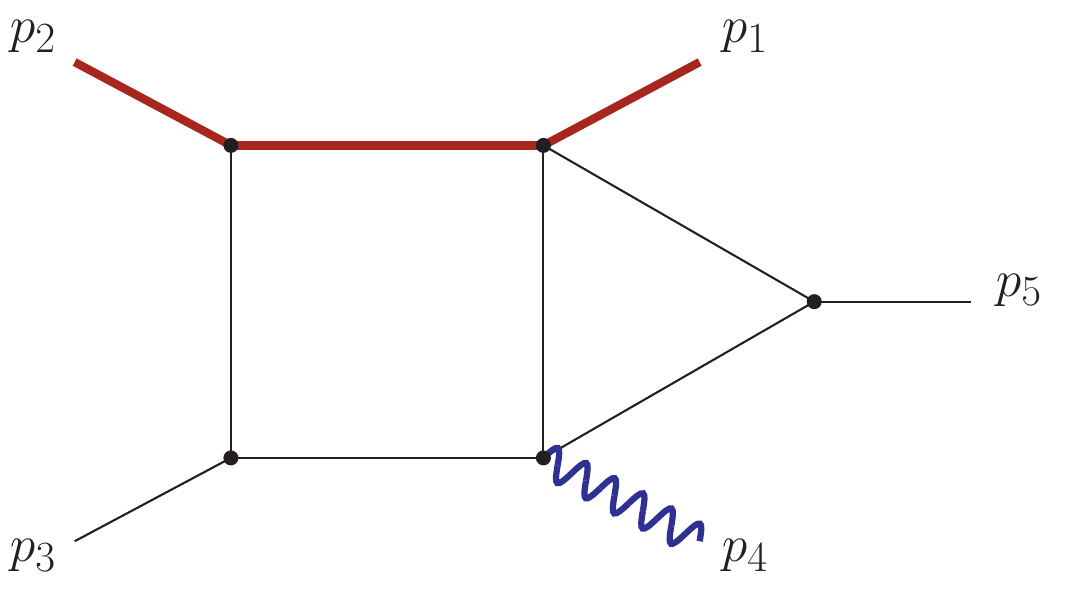}
  \end{figure}
\end{minipage}%
\hspace{0.02\textwidth}
\begin{minipage}{0.61\textwidth}
  $\begin{aligned}
     N_{31}^{(F_{2})}& = \eps^3 r_1 \frac{\mu_{11}}{D_{F_2,8}} \,, \\
     N_{32}^{(F_{2})}& = \eps^4 \left(s_{23}-m_t^2\right) (s_{34}-s_{12}) \,.
  \end{aligned}$
\end{minipage}

\end{samepage}

\medskip

\begin{samepage}

\begin{center}
  \textbf{$F_{2}$, $(0,1,1,0,1,1,1,1,0,0,0)$, 5 MIs}
\end{center}
\hspace{\parindent}
\begin{minipage}{0.37\textwidth}
  \begin{figure}[H]
    \centering
    \includegraphics[width=\textwidth]{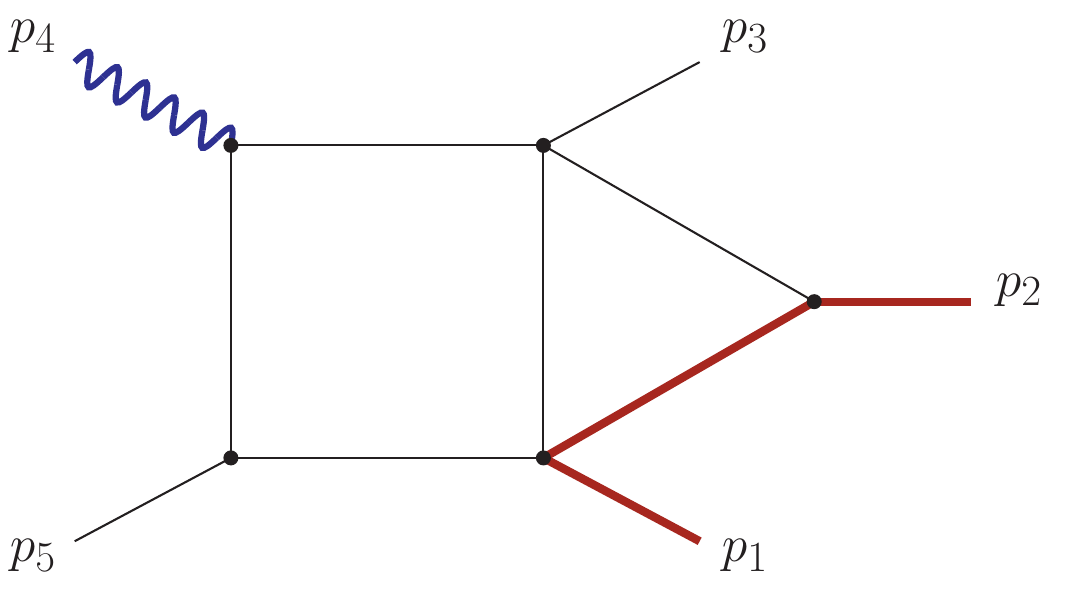}
  \end{figure}
\end{minipage}%
\hspace{0.02\textwidth}
\begin{minipage}{0.61\textwidth}
  $\begin{aligned}
    N_{24}^{(F_{2})}&= \eps^3 r_{1} \frac{\mu_{22}}{D_{F_2,8}} \,, \\
    N_{25}^{(F_{2})}&= \eps^3 r_{1} \frac{\mu_{12}}{D_{F_2,8}} \,, \\
    N_{26}^{(F_{2})}&= \eps^4 \left[ m_w^2 s_{12} - \left(m_t^2 - s_{15} + s_{34}\right) s_{45}\right] \\
      & \phantom{=} \ + \eps^3 m_t^2 \frac{m_w^2 s_{12} - s_{34} s_{45}}{D_{F_2,2}} \,, \\
    N_{27}^{(F_{2})}&= \eps^4 \left[m_w^2 s_{12} - (m_t^2 - s_{15} + s_{34}) s_{45}\right] \\
      & \phantom{=} \ + \eps^3 m_t^2 \frac{m_w^2 D_{F_2,1} - s_{45} D_{F_2,9}}{D_{F_2,2}} \\
      & \phantom{=} \ + \text{(sub-sectors)} \,, \\
    N_{28}^{(F_{2})}&= \eps^4 r_{14} \,.
  \end{aligned}$
\end{minipage}

\end{samepage}

\medskip

\begin{samepage}

\begin{center}
  \textbf{$F_{2}$, $(1,1,1,1,0,1,0,1,0,0,0)$, 2 MIs}
\end{center}
\vspace{-0.5cm}
\hspace{\parindent}
\begin{minipage}{0.37\textwidth}
  \begin{figure}[H]
    \centering
    \includegraphics[width=\textwidth]{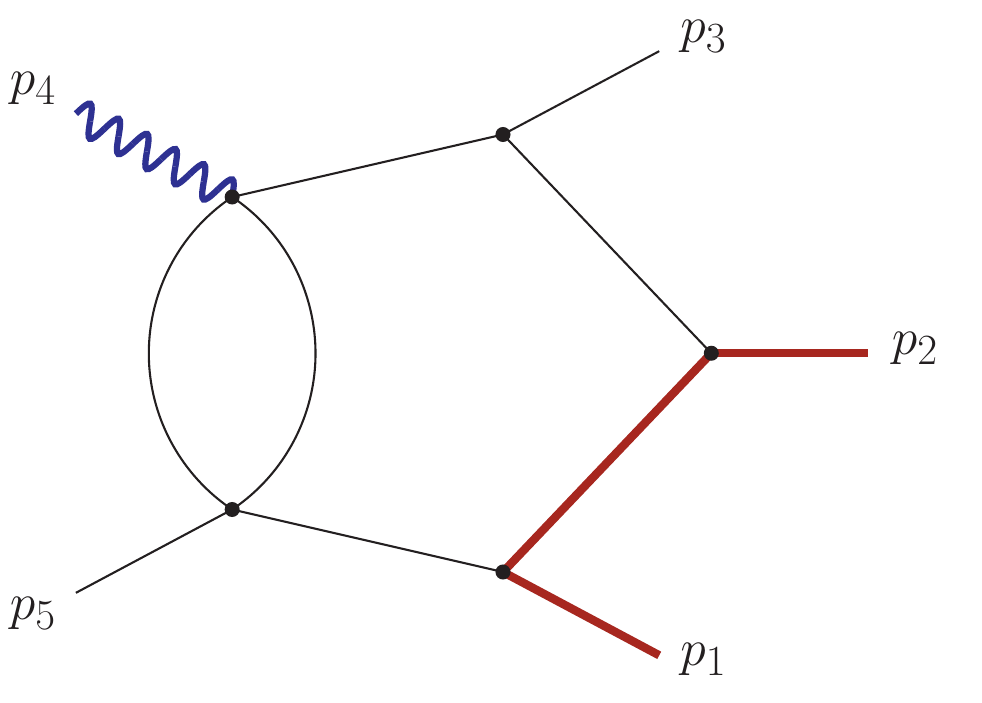}
  \end{figure}
\end{minipage}%
\hspace{0.02\textwidth}
\begin{minipage}{0.61\textwidth}
  $\begin{aligned}
    N_{29}^{(F_{2})}&= \eps^3 r_1 \frac{\mu_{11}}{D_{F_2,8}} \,, \\
    N_{30}^{(F_{2})}&= \eps^3 (2\eps-1) s_{12} \left(m_t^2 - s_{23}\right) \,.
  \end{aligned}$
\end{minipage}

\end{samepage}

\medskip

\begin{samepage}

\begin{center}
  \textbf{$F_{3}$, $(1,1,1,1,0,1,0,1,0,0,0)$, 3 MIs}
\end{center}
\vspace{-0.5cm}
\hspace{\parindent}
\begin{minipage}{0.37\textwidth}
  \begin{figure}[H]
    \centering
    \includegraphics[width=\textwidth]{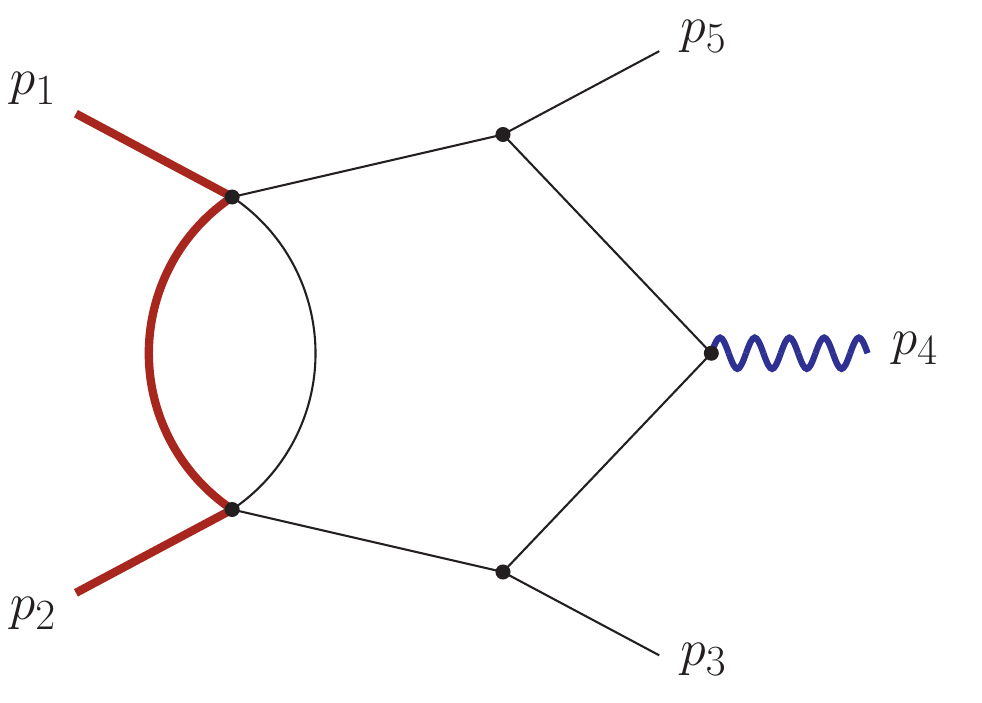}
  \end{figure}
\end{minipage}%
\hspace{0.02\textwidth}
\begin{minipage}{0.61\textwidth}
  $\begin{aligned}
    N_{37}^{(F_{3})}&= \eps^3 \left(m_w^2 s_{12} - s_{34} s_{45}\right) \frac{m_t^2 + D_{F_3,9}}{D_{F_3,6}} \,, \\
    N_{38}^{(F_{3})}&= \eps^3 r_1 \frac{\mu_{11}}{D_{F_3,6}} \,, \\
    N_{39}^{(F_{3})}&= \eps^3 r_1 \frac{\mu_{11}}{D_{F_3,8}} \,.
  \end{aligned}$
\end{minipage}

\end{samepage}

\medskip

\begin{samepage}

\begin{center}
  \textbf{$F_{1}$, $(1,1,1,1,0,1,0,1,0,0,0)$, 2 MIs}
\end{center}
\vspace{-0.5cm}
\hspace{\parindent}
\begin{minipage}{0.37\textwidth}
  \begin{figure}[H]
    \centering
    \includegraphics[width=\textwidth]{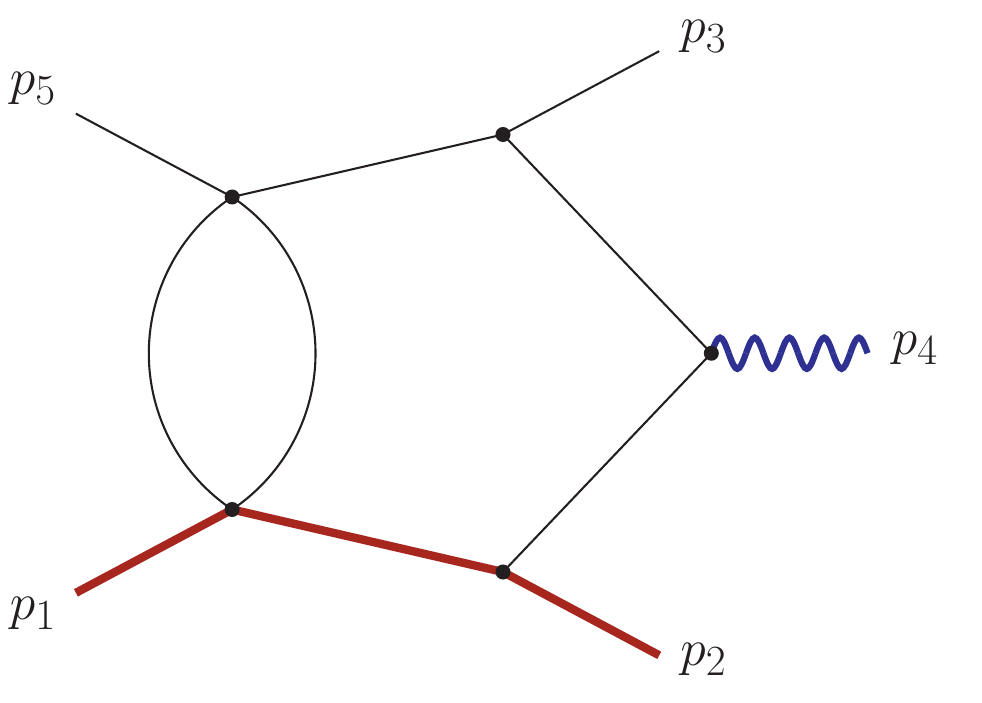}
  \end{figure}
\end{minipage}%
\hspace{0.02\textwidth}
\begin{minipage}{0.61\textwidth}
  $\begin{aligned}
    N_{29}^{(F_{1})}&= \eps^3 r_1 \mu_{11}/ D_{F_1,8} \,, \\
    N_{30}^{(F_{1})}&= \eps^2 (1-2\eps) s_{34}(m_t^2-s_{23}) \,. \\
  \end{aligned}$
\end{minipage}

\end{samepage}

\section{Details of the elliptic curves}
\label{app:elliptic_details}

In this appendix, we summarise our analysis of the polynomial structures
associated with the five-point elliptic sector presented
in~\cref{sec:five-point_elliptic}.
We begin with a generic quartic polynomial defining the elliptic curve:
\begin{align}
    \cP\brk{z}
    =
    c_4 \, z^4 + c_3 \, z^3 + c_2 \, z^2 + c_1 \, z + c_0
    \label{eq:elliptic_quartic}
    \>.
\end{align}
In the case of the polynomial $\mathcal{P}_{5\text{-pt.}}$
appearing in \cref{eq:elliptic_monster},
the coefficients $c_i = c_i\brk{\vec{x}, r_1}$ are polynomials in the kinematic
variables $\vec{x}$ and the square root $r_1$ defined in~\cref{eq:roots1}.
Under the action of the M\"o{}bius transformation
\begin{align}
    z \mapsto \frac{a_1 + a_2 \, \zt}{a_3 + a_4 \, \zt} \>,
    \label{eq:moebius_transformation}
\end{align}
the elliptic integration kernel transforms as
\begin{align}
    \frac{\dd z}{\sqrt{\cP\brk{z}}}
    \mapsto
    \brk{a_2 \, a_3 - a_1 \, a_4} \, \frac{\dd \zt}{\sqrt{\widetilde{\cP}\brk{\zt}}}
    \>,
    \label{eq:moebius_transformation_integrand}
\end{align}
leading to a new quartic polynomial $\widetilde{\cP}$, which takes the form
\begin{align}
    \widetilde{\cP}\brk{\zt}
    =
    \ct_4 \, \zt^4 + \ct_3 \, \zt^3 + \ct_2 \, \zt^2 + \ct_1 \, \zt + \ct_0
    \>.
\end{align}
This transformation provides sufficient freedom to bring the defining polynomial into
Weierstra\ss{} form:
\begin{align}
    \widetilde{\cP}^{\text{W.}}\brk{\zt} = 4 \, \zt^3 + g_1 \, \zt + g_0
    \label{eq:elliptic_weierstrass}
    \>.
\end{align}
To determine the four transformation parameters $a_i$
in~\cref{eq:moebius_transformation}, we impose the following four conditions:
the $\zt$-independent prefactor on the RHS of~\cref{eq:moebius_transformation_integrand}
must be unit, the coefficients $\ct_4$ and $\ct_2$ must vanish, and
the leading coefficient must be normalised to $\ct_3 = 4$.
These conditions are enforced by constructing a Gr\"obner basis of the corresponding
ideal and performing polynomial division on the two coefficients
in~\cref{eq:elliptic_weierstrass} ($g_0$ and $g_1$) w.r.t.\ to this basis. As a result, the
coefficients in~\cref{eq:elliptic_weierstrass} are expressed in
terms of those in~\cref{eq:elliptic_quartic} as follows:
\begin{align}
  \begin{aligned}
    g_1 &= \frac{
        - 12 \, c_0 \, c_4
        + 3 \, c_1 \, c_3
        - c_2^2
    }{
        12
    }
    \>,
    \\
    g_0 &= \frac{
        - 72 \, c_0 \, c_2 \, c_4
        + 27 \, c_0 \, c_3^2
        + 27 \, c_1^2 \, c_4
        - 9 \, c_1 \, c_2 \, c_3
        + 2 \, c_2^3
    }{
        432
    }
    \>.
  \end{aligned}
    \label{eq:elliptic_weierstrass_coefficients}
\end{align}
A crucial observation is that, after substituting
the coefficients $c_i = c_i\brk{\vec{x}, r_1}$
of the polynomial $\mathcal{P}_{5\text{-pt.}}$
into \cref{eq:elliptic_weierstrass_coefficients}, the square root $r_1$
cancels out in the coefficients $g_0 = g_0\brk{\vec{x}}$ and
$g_1 = g_1\brk{\vec{x}}$, resulting in the same Weierstra\ss{} polynomial,
$\cP^{\text{W.}}_{5\text{-pt.}}$, for both $\mathcal{P}_{5\text{-pt.}}$
and its $r_1$-conjugate $\mathcal{P}_{5\text{-pt.}}^{\dagger}$.
The discriminant of $\cP^{\text{W.}}_{5\text{-pt.}}$ takes the form
\begin{align}
  \begin{aligned}
    \Delta_{E_{5\text{-pt.}}}
      & \propto g_1^3 + 27 \, g_0^2 \\
      & \propto m_t^2 \, m_w^2 \, s_{12}^6 \, \bigbrk{G\brk{p_2, p_3}}^{6}
    \> \bigbrk{G\brk{p_1, p_2, p_3}}^{12}
    \> y_{307}
    \>,
  \end{aligned}
    \label{eq:elliptic_monster_discriminant}
\end{align}
where we have made use of the Gram determinants defined
in~\cref{eq:gram-definitions} to obtain a more compact representation.
The factor $y_{307}$\footnote{This notation makes contact with the symbol used in the ancillary files~\cite{zenodo}.}
in~\cref{eq:elliptic_monster_discriminant} is the irreducible degree-14
polynomial in $\vec{x}$ introduced in~\cref{sec:five-point_elliptic}, which
appears among the singularities of the connection matrix.

From the structure of the coefficients
in~\cref{eq:elliptic_weierstrass_coefficients}, it follows naturally that the
$j$-invariants of the curves in~\cref{eq:5pt_ell_curves} are purely rational functions,
\begin{align} \label{eq:jinv_5pt}
  \begin{aligned}
    j_{5\text{-pt.}}
    & = 1728 \, \frac{g_1^3}{g_1^3 + 27 \, g_0^2} \\
    & = \frac{
        Q(\vec{x})^3
    }{
        m_t^4 \, m_w^4 \,
        y_{307}
    }
    \>,
  \end{aligned}
\end{align}
where $Q$ is a degree-6 polynomial in $\vec{x}$ whose precise form is irrelevant to our discussion.
The equality of the $j$-invariants\footnote{
    We also verified this equality numerically from the original quartic
    representation in~\cref{eq:elliptic_quartic}.
} of the elliptic curves defined by
$\mathcal{P}_{5\text{-pt.}}$ and $\mathcal{P}_{5\text{-pt.}}^{\dagger}$ proves
the uniqueness of the elliptic curve for the five-point sector studied in
\cref{sec:five-point_elliptic}.

\bibliographystyle{JHEP}
\bibliography{biblio}

\end{document}